%% file: main.tex
\documentclass[sigconf]{acmart}
\makeatletter
\renewcommand\@formatdoi[1]{\ignorespaces}
\makeatother

\setcopyright{none}

\renewcommand\footnotetextcopyrightpermission[1]{}

\acmConference[]{}{}{}
\acmDOI{}
\acmISBN{}
\acmYear{}
\copyrightyear{}
\acmPrice{}
\acmSubmissionID{}

\usepackage{graphicx} 
\usepackage{booktabs} 
\usepackage{fixme}
\usepackage{float}
\usepackage{stfloats}
\usepackage{booktabs}
\usepackage{multirow}
\usepackage{graphicx}
\usepackage{subcaption}
\usepackage{calc}
\usepackage{xcolor}
\usepackage{tcolorbox}
\tcbuselibrary{breakable}
\usepackage{quoting}
\usepackage{caption}

\author{Shayan Talaei}
\affiliation{
  \institution{Department of Management Science and Engineering, Stanford University}
  \city{Stanford}
  \state{CA}
  \country{USA}
}
\email{stalaei@stanford.edu}

\author{Meijin Li}
\affiliation{
  \institution{Department of Electrical Engineering, Stanford University}
  \city{Stanford}
  \state{CA}
  \country{USA}
}
\email{meijin@stanford.edu}

\author{Kanu Grover}
\affiliation{
  \institution{Department of Computer Science, Stanford University}
  \city{Stanford}
  \state{CA}
  \country{USA}
}
\email{groverk@stanford.edu}

\author{James Kent Hippler}
\affiliation{
  \institution{Department of Management Science and Engineering, Stanford University}
  \city{Stanford}
  \state{CA}
  \country{USA}
}
\email{khippler@stanford.edu}

\author{Diyi Yang}
\affiliation{
  \institution{Department of Computer Science, Stanford University}
  \city{Stanford}
  \state{CA}
  \country{USA}
}
\email{diyiy@cs.stanford.edu}

\author{Amin Saberi}
\affiliation{
  \institution{Department of Management Science and Engineering, Stanford University}
  \city{Stanford}
  \state{CA}
  \country{USA}
}
\email{saberi@stanford.edu}

\begin{document}

\title{StorySage: Conversational Autobiography Writing Powered by a Multi-Agent Framework}

\begin{abstract}
Every individual carries a unique and personal life story shaped by their memories and experiences. However, these memories are often scattered and difficult to organize into a coherent narrative—a challenge that defines the task of autobiography writing. Existing conversational writing assistants tend to rely on generic user interactions and pre-defined guidelines, making it difficult for these systems to capture personal memories and develop a complete biography over time. We introduce \textit{StorySage}, a user-driven software system designed to meet the needs of a diverse group of users that supports a flexible conversation and a structured approach to autobiography writing. Powered by a multi-agent framework composed of an Interviewer, Session Scribe, Planner, Section Writer, and Session Coordinator, our system iteratively collects user memories, updates their autobiography, and plans for future conversations. In experimental simulations, \textit{StorySage} demonstrates its ability to navigate multiple sessions and capture user memories across many conversations. User studies ($N=28$) highlight how \textit{StorySage} maintains improved conversational flow, narrative completeness, and higher user satisfaction when compared to a baseline. In summary, \textit{StorySage} contributes both a novel architecture for autobiography writing and insights into how multi-agent systems can enhance human-AI creative partnerships.
\end{abstract}



\begin{teaserfigure}
\centering
\includegraphics[width=0.98\textwidth]{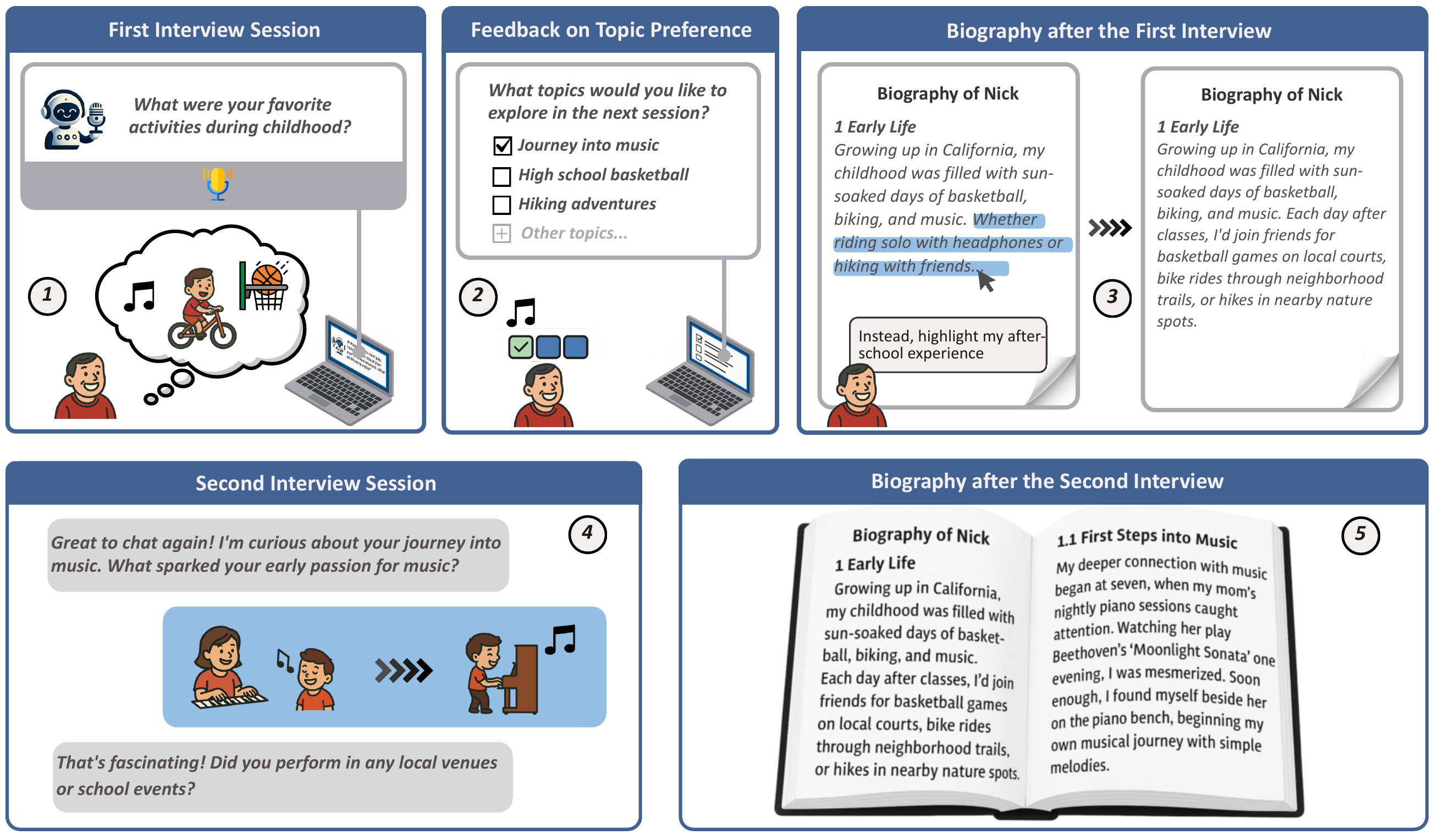}
\caption{\textbf{User interaction flow with \emph{StorySage}, as illustrated through our hypothetical user Nick.} \textbf{(1)} Nick has a conversation with \emph{StorySage} where he shares his childhood memories. \textbf{(2)} After the interview, Nick indicates he wants to talk about his \textit{journey into music} during the next session. \textbf{(3)} Nick receives his updated autobiography after the first session. He chooses to provide suggestions to \textit{StorySage} to emphasize his after-school experience, and receives his updated autobiography. \textbf{(4)} Nick begins a second interview, picking up where he left off in the previous conversation to begin discussing about his journey into music. After concluding the interview and selecting discussion topics for next session, \textbf{(5)} Nick can review his updated biography with the newly added story.}
\label{fig:product_workflow}
\end{teaserfigure}

\maketitle
\pagestyle{plain} 

\begin{teaserfigure}
\centering
\includegraphics[width=\textwidth]{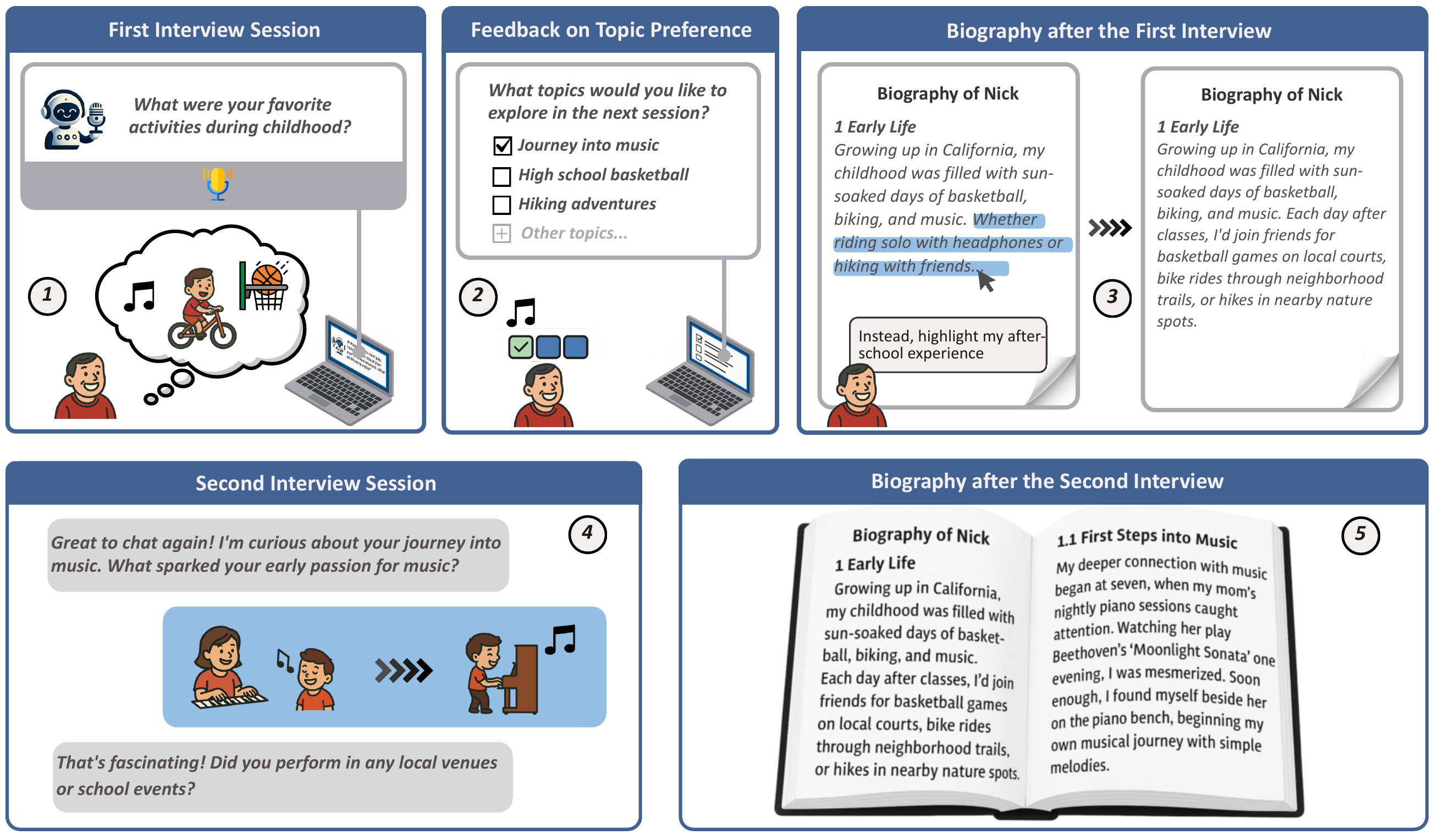}
\caption{\textbf{User interaction flow with \emph{StorySage} illustrated through Nick's experience.} (1) Nick engages in a conversation with \emph{StorySage} sharing his childhood memories. (2) After the interview, Nick selects ``Journey into music" as a topic to discuss in the next session, which will be included as a part of the next interview agenda. (3) Nick can review and edit his updated biography either directly or by leaving natural language comments for the \emph{StorySage} to do that. The second interaction of Nick starts at (4), where he engages in a deeper conversation on his memories on musical journey. Similarly, after finishing the interview and selecting his topics of interests, (5) Nick can review his updated biography with the updated subsections.}
\label{fig:product_workflow}
\end{teaserfigure}



\section{Introduction}

Life narratives are deeply unique and valuable ~\citep{mcadams2013narrative, fivush2011autobiographical, bluck2000lifestory}, shaped through an interplay of personal memories that involve achievements, struggles, and moments of reflection. Capturing and expressing these narratives through an autobiography is a powerful way for people to preserve their legacy~\citep{butler1963life, allen2008legacy, cuevas2021dignity, hunter2005leaving}, deepen their self-understanding~\citep{pennebaker1999forming, mclean2007selves}, and share their life journey with others, which in turn, fosters a strong connection across generations and communities~\citep{alea2003you, alea2007ll, fivush2011personal}. 

Autobiography writing presents a few unique challenges. It is a deeply personal process that involves recalling forgotten memories, which can be emotionally demanding and time-consuming~\citep{conway2000selfmemory, linde1993lifestories, schacter1999sevensins, williams2007autobiographical}. Additionally, it relies on long-form storytelling, whereas most writing systems are designed for short-term content. For example, MindScape~\citep{nepal2024contextual} supports context-aware journaling but focuses on short-term self-reflection. Recently, GuideLLM~\citep{duan2025guidellmexploringllmguidedconversation} introduced an autobiography writing system where a system autonomously guides users through structured interview sessions. Their approach draws on conversational guidelines from "The Life Story Interview"~\citep{mcadams2008life}, with each session focused on a single topic, used to create a chapter of the user’s autobiography.

To understand the effectiveness of these existing frameworks, we consulted with experts and conducted a focus group study. These highlighted the importance of designing a system that can hold conversations that feel engaging and personal. Furthermore, the system should sufficiently capture memories shared by users while preserving their factual accuracy in writing. Most importantly, users expressed a strong preference for systems that allow them to control the narrative direction and writing process; an observation also echoed in previous work~\citep{gero2023socialdynamics, behrooz2024holdinglinestudywriters, kim2024diarymate, li2024valuebenefitsconcernsgenerative}.

In this work, we present a novel framework that is \emph{StorySage}\footnote{We open-source our code, including implementations of the agents and their prompts, available at \href{https://github.com/ShayanTalaei/AI_friend_biographer}{this link}.}, a conversational autobiography writing software that attempts to both \textit{guide} users in recalling and organizing their memories, while also actively \textit{involving} them in the storytelling and writing process. The system is meant to support human-AI co-creativity, and align with the key design goals we identified from our formative study. Although existing work in this space is intuitive and easy to use, our approach offers a new perspective on autobiography writing that emphasizes flexibility and collaboration.

Users interact with \textit{StorySage} over the course of multiple sessions. During each session, users engage in a flexible-length, dynamic conversation with the system that feels personal to their interests. Following each session, users provide a list of discussion topics to explore in future conversation. They also receive an updated autobiography after every session, which they can review and edit prior to the next session. In the backend, this workflow is managed by a multi-agent framework composed of five specialized agents. The Interviewer and Session Scribe are responsible for facilitating a natural, responsive conversation for the user, while the Planner and Section Writer together outline and incorporate users' memories into their autobiography. Lastly, the Session Coordinator oversees continuity by preparing a guiding agenda for future sessions.

\textit{StorySage} is a software system built for a general population of users that are interested in writing their autobiography, reflective of the demographic diversity among participants in our user study. We evaluate the effectiveness of \textit{StorySage} through a simulation-based experiment and user study with $N=28$ participants by comparing it against a \textit{Baseline}. We then present qualitative findings that provide deeper insight into user experience across both systems. In summary, this paper presents two main contributions: the introduction of \textit{StorySage}, a novel user-centric system for conversational autobiography writing that supports human-AI co-creativity, and an evaluation of its effectiveness through a real-world user study.

\section{Related Work}
\subsection{Autobiographical Memory and Reflection}

Autobiographical memory, or memory about one's past, plays an important role in how people form their identity, connect with others, and make decisions about their future \citep{conway2000selfmemory, habermas2000getting, mcadams2001psychology, cuevas2021dignity}. Studies in psychology indicate that recalling and documenting these memories can improve mental health~\citep{hallford2024guided, valtonen2022health}, strengthen relationships~\citep{moscovitch2024hooking, speer2020social}, and improve memory function over time~\citep{serrano2004life, subramaniam2014life}.

Motivated by the psychological benefits of memory elicitation, many digital tools have emerged—from early “lifelogging” visions \citep{gummell2006mylifebits} to systems for journaling, reminiscence, and narrative self-reflection~\citep{peesapati2010pensieve, tucci2024retromind, kim2024mindfuldiary, white2023memory, elsden2016s, kim2024mindfuldiary}. While methods like journaling and memoir writing help individuals organize memories, they often fall short in prompting recall and supporting deeper reflection \citep{ma2017writeforlife}. To address these limitations, systems like Pensieve~\citep{peesapati2010pensieve} use social media-based prompts to enhance daily reminiscence. RetroMind~\citep{tucci2024retromind} combines conversational and visual cues for reminiscence therapy in dementia care, and SimSensei~\citep{morbini2014simsensei} leverages virtual conversational agents for storytelling. Reflective tools like MindfulDiary~\citep{kim2024mindfuldiary} build on this work by using conversational agents to support journaling in mental health contexts \citep{ma2017writeforlife}.

To support memory retrieval and deeper reflection, researchers have examined interviewing strategies for autobiography writing. Harding et al.~\citep{biographicalinterviewing2006} describe two primary approaches: chronological interviews, which elicit life events in sequence, and narrative-focused interviews, which explore clusters of meaningful experiences in detail. They advocate for a hybrid interviewing style that blends narrative and explanatory elements to encourage users to explore memories they find meaningful. Jiménez and Orozco~\cite{jimenez2021prompts} further propose interviewing techniques that include “grand tour” questions, “counterfactuals,” and “comparisons” to encourage interviewees to reflect more deeply on their experiences.

\subsection{Human–AI Co-Creation}

Recent advances in artificial intelligence (AI) have significantly expanded the capabilities of AI-assisted writing tools, transforming how users engage with writing tasks across academic, professional, and everyday contexts~\citep{Lee_2024}. These tools include predictive text suggestions that enhance  productivity~\citep{chen2019gmailsmartcomposerealtime, arnold20:predictcive, CambonEarlyLT}, interactive systems that support complex editing and revision workflows~\citep{Afrin_2021, faltings-etal-2021-text, Dang_2022, karolus2023readability, cotos2020rwt, knight2020acawriter, kim2023explainableaiwritingassistants}, and even collaborative assistants that facilitate human-AI creative partnerships~\citep{goldfarb-tarrant-etal-2019-plan, Biermann2022FromTT, chakrabarty2022helpwritepoeminstruction}. These writing assistants have been applied in scientific and academic writing~\citep{boillos2025aicollaboration, shao2024assistingwritingwikipedialikearticles, shen2023convxaideliveringheterogeneousai, gero2021sparksinspirationsciencewriting}, personal and creative expression~\citep{nepal2024contextual, swanson-etal-2021-story, li2024diaryhelperexploringuseautomatic}, and professional business communication~\citep{chen2019gmailsmartcomposerealtime, de2023formalstyler, omelianchuk-etal-2020-gector}. 

Moving beyond general-purpose writing support, many tools have been designed for creative narrative writing~\citep{coenen2021wordcrafthumanaicollaborativeeditor, Zhang_2022, radwan2024sardhumanaicollaborativestory, swanson2012sayanything}. These systems facilitate collaboration between humans and AI to craft fictional stories, scripts, poems, and other forms of content. For example, Story Centaur~\citep{swanson-etal-2021-story} is an interactive platform that leverages few-shot prompt engineering to enable writers to create customized narratives, while Wordcraft~\citep{coenen2021wordcrafthumanaicollaborativeeditor} is a collaborative storytelling editor that offers writers control over story continuations and stylistic edits. Another such system, Dramatron~\citep{mirowski2023dramatron}, is a screenplay and playwriting assistant that employs hierarchical prompting techniques to iteratively generate narrative content.

While AI-assisted writing tools offer significant benefit, recent empirical studies have raised concerns about their impact on authorship, voice, and creativity~\citep{gero2023socialdynamics, behrooz2024holdinglinestudywriters, kim2024diarymate, kim2024authors}. For instance, \citet{behrooz2024holdinglinestudywriters} found that while writers may be open to receiving AI support, they stress the need for clear boundaries to maintain creative agency. Similarly, \citet{li2024valuebenefitsconcernsgenerative} found that while AI assistance can enhance productivity and confidence, they may also reduce authors' sense of ownership and reduce diversity in writing style. These findings highlight the importance of preserving human agency in writing to ensure that writers retain voice and style in their work.

Within this context, \emph{StorySage} positions itself as a human-AI collaborative system intended for autobiographical writing.  \emph{StorySage} is designed to preserve human agency by enabling flexible story navigation and adapting the conversation and narrative in response to user feedback, addressing the concerns raised by \citep{gero2021sparksinspirationsciencewriting, behrooz2024holdinglinestudywriters, li2024valuebenefitsconcernsgenerative, kim2024diarymate}.

\subsection{LLM-Powered Multi-Agent Systems}
Multi-agent systems (MAS) leverage a set of specialized agents that interact to solve complex tasks, by adopting distinct roles, capabilities, and communication protocols \citep{wooldridge2009multiagent}. With the advent of large language models (LLMs), researchers have begun to design LLM-powered multi-agent systems in domains like writing \citep{lee2025map, shao2024assistingwritingwikipedialikearticles}, coding \citep{hong2024metagptmetaprogrammingmultiagent, qian2024chatdevcommunicativeagentssoftware, wu2023autogenenablingnextgenllm, talaei2024chesscontextualharnessingefficient}, and social simulation \citep{park2023generative, chen2023agentversefacilitatingmultiagentcollaboration, zhou2024sotopiainteractiveevaluationsocial}. 

Accomplishing complex tasks with an LLM-powered multi-agent framework necessitates careful orchestration \citep{tran2025multiagentcollaborationmechanismssurvey}. Prior studies have proposed different approaches to facilitate this coordination, ranging from predefined sequential interactions~\citep{hong2024metagptmetaprogrammingmultiagent, zhang2024chainagentslargelanguage} to more complex communication patterns~\citep{wang2024unleashingemergentcognitivesynergy, du2023improvingfactualityreasoninglanguage, li2025parallelizedplanningactingefficientllmbased}. Other work \citep{fourney2024magenticonegeneralistmultiagentsolving, shen2023hugginggptsolvingaitasks} has incorporated a central orchestrator responsible for planning and outlining tasks, with subordinate agents dedicated to executing these predefined plans. This approach has demonstrated a good balance between simplicity and successful task completion. 

Moreover, having an effective memory mechanism is essential for enabling long-term goal navigation in multi-agent systems \citep{zhang2024surveymemorymechanismlarge, georgeff1999bdi}. For instance, Generative Agents \citep{park2023generative} introduced complex memory architectures capable of capturing, summarizing, and reflecting on extensive individual experiences, thus enabling agents to plan and interact over prolonged periods. Additionally, the LLaMAC framework \citep{zhang2024controllinglargelanguagemodelbased} addresses the challenges of hallucination and coordination in large-scale multi-agent systems by incorporating actor-critic mechanisms inspired by value distribution encoding in the human brain. At the architecture level, the Long-Short Decision Transformer (LSDT) \citep{wang2025longshort} integrates parallel self-attention and convolutional branches to effectively capture both global and local dependencies in sequential decision-making tasks. In the context of human-AI interactions, robust memory modules are particularly crucial for delivering personalized and contextually relevant discussions \citep{serban2016buildingendtoenddialoguesystems, zhang-etal-2018-personalizing, gummell2006mylifebits}. Similarly, OmniQuery~\citep{li2025omniquerycontextuallyaugmentingcaptured} and MAP~\citep{lee2025map} apply RAG techniques to retrieve relevant memories from a user's personal archive in order to provide a personalized user experience.

Building on these practices, \emph{StorySage} introduces a multi-agent architecture designed for autobiographical interviewing and writing. The system orchestrates five specialized agents alongside a persistent memory module to address key challenges in MAS design, including maintaining narrative coherence, adapting to personal context, and sustaining user engagement~\citep{wu2023autogenenablingnextgenllm, park2023generative, lee2025map, zhang2018personalizingdialogueagentsi}.

\section{Design of StorySage}

\subsection{Formative Study}

To inform the design of \textit{StorySage}, we conducted an initial pilot study\footnote{This pilot study was exploratory in nature. To remain consistent with IRB guidelines for non-research pilot activities, we do not report statistical findings or identifiable participant information from preliminary interviews.} consisting of ten 60-minute conversations with various individuals—a professional biographer, technology strategist, and eight users interested in biography writing. Drawing from principles in the Design Thinking Bootleg~\cite{dschoolbootleg}, we explored how people engage in conversations about their memories, their perceptions of AI, and challenges they anticipate in writing their autobiography. These early discussions with experts emphasized the importance of building trust between interviewer and interviewee by asking both personal, rapport-building and reflective questions, which echoes insights from prior work in biographical interviewing techniques ~\cite{biographicalinterviewing2006, jimenez2021prompts}. These conversations also highlighted the importance of gathering feedback on early versions of the system to better understand user needs and expectations. 

Drawing from prior work in AI-assisted autobiography writing, namely GuideLLM~\citep{duan2025guidellmexploringllmguidedconversation}, we then designed a prototype with a fixed conversational agenda and writing style, which we tested through an interactive focus group study. Our focus group consisted of 20 senior citizens, primarily chosen for their rich life experiences and interest in memoir writing. Participants were invited to hold a natural, 10-minute conversation with the prototype which was described as a system that would ask questions and produce a draft of their autobiography. However, participants found that the system's conversational ability lacked depth and felt impersonal. One noted, “this doesn’t feel like a conversation. It feels like a series of one-sided questions.” A subset of individuals familiar with ChatGPT agreed that they "would rather provide a set of conversational topics to the chatbot and receive questions to answer." Many participants had previously worked with professional memoirists and described the appeal in writing as a form of mental stimulation; following their conversation with the prototype, they perceived the system to be “a lazy way to write” and expressed a desire for greater ownership. Additionally, one participant observed "a hallucination in sections where [I provided] limited information," while another felt that the output “doesn’t sound like my story, more like a list of experiences.” These insights from our formative study guided several key design aspects of \textit{StorySage}, which we outline below.

\subsection{Design Goals}\label{sec:design_goals}

\begin{figure*}[t]
  \centering
  \includegraphics[width=\textwidth]{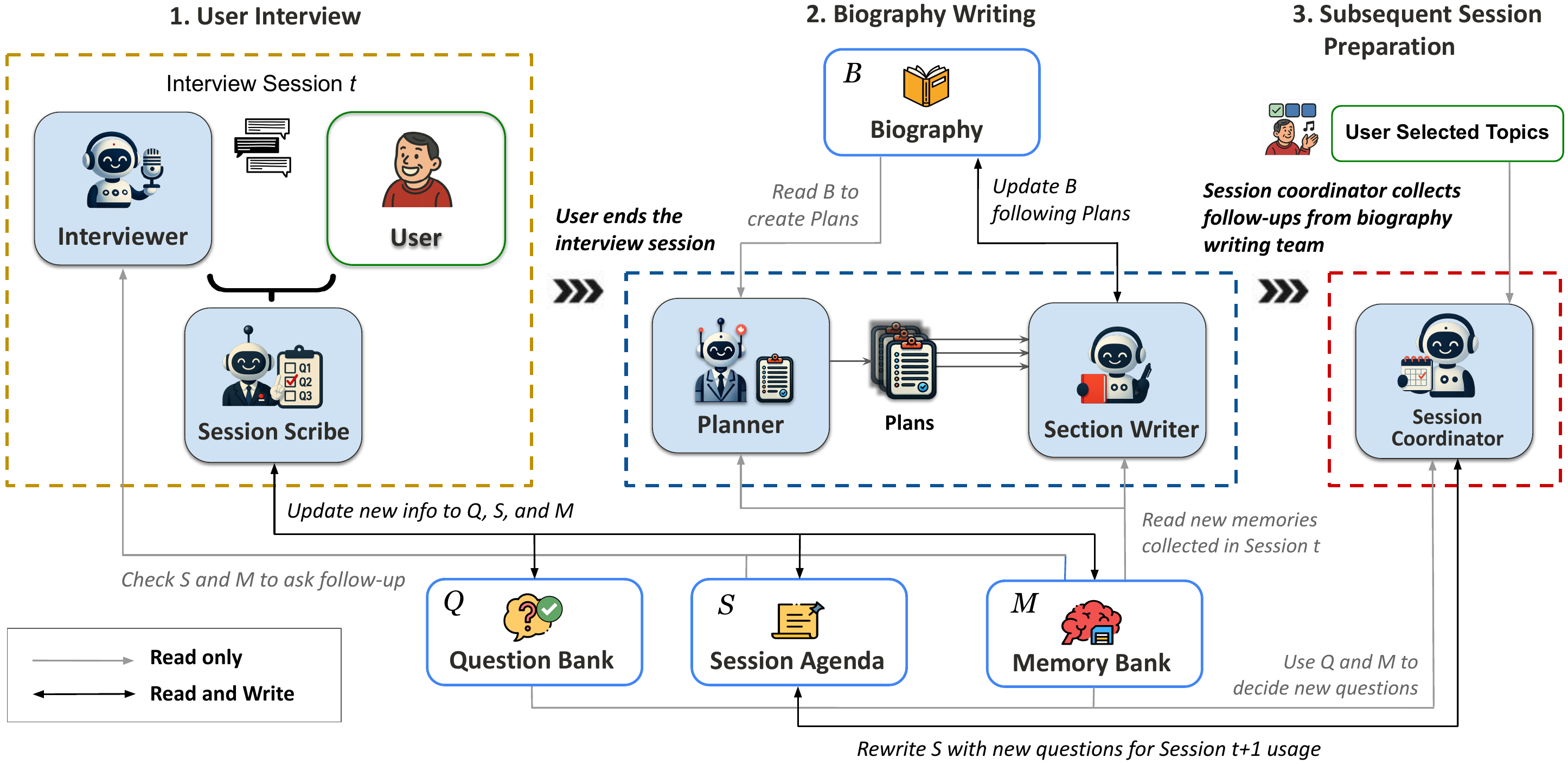}
  \caption{\textbf{Overview of the \emph{StorySage} multi-agent architecture.} (\textbf{User Interview}) In this phase, the Interviewer engages the user in a conversation to help them share their memories, while the Session Scribe works in the background to log key details from the conversation—including answered questions and shared memories—and generates follow-ups. \textbf{(Biography Writing)} After the user concludes the interview, the Planner analyzes the existing biography structure and new memories from the conversation to formulate a set of structured update plans. Subsequently, the Section Writer uses these plans to write narrative content in the user's autobiography. \textbf{(Subsequent Session Preparation)} Finally, the Session Coordinator begins planning for the next session by designing a session agenda.}
  \label{fig:architecture_overview}
\end{figure*}

\label{goal:1_human_agency}

\paragraph{\textbf{DG 1: Providing Human Agency}} Aligned with expert advice to prioritize user-led conversations, feedback from our focus group study revealed that users wanted more control over the conversational flow and sought involvement in the writing process as a form of mental stimulation and a way to personalize their story. These insights suggest that an effective system should adopt a human-AI co-creative approach by providing interactive workflows that allow users to guide the conversation and shape the narrative, while also being designed to adapt the conversational flow and narrative structure in real time.

\paragraph{\label{goal:3_conv} \textbf{DG 2: Fostering Natural Conversation}} An effective system should be able to maintain a natural conversation flow by asking thoughtful questions relevant to the users' interests. Insights from our focus group study revealed that some users found the prototype’s conversational template overly one-sided, with many prompts being abstract in nature. This aligns with concerns raised in our literature review, which cautions against relying solely on surface-level questions~\citep{biographicalinterviewing2006}—though they can be effective as guiding prompts. These findings point to the need for balance between guiding prompts and more thought-provoking follow-up questions that support memory recall. Experts similarly emphasize that conversations which begin by building familiarity and trust naturally lead individuals to reflect and share deeper stories. Additionally, fostering more meaningful conversations requires the system to recall prior interactions and sustain conversational continuity across multiple sessions.

\paragraph{\label{goal:4_integrity} \textbf{DG 3: Preserving Narrative Integrity}} A recurring theme in our early exploratory interviews and focus group study was mistrust in AI handling of personal memories, particularly due to the presence of hallucinations. These concerns reflect a broader expectation: when users share their life stories, they expect the system not only to listen, but also to remember and accurately reflect the story in their autobiography. To this point, an autobiography writing system should accurately represent the user's memories and adequately capture the content they share in their conversation with an appropriate level of detail. Building on DG 1, such a system should also provide users with mechanisms to incorporate their voice and correct inaccuracies in the autobiography, ensuring that narrative integrity is preserved.

\paragraph{\textbf{DG 4: Maintaining Responsiveness.}} Noting that some focus group participants felt that long response times disrupted conversation flow, we believe a well-designed system should both propose follow-up questions and generate the autobiography in a timely manner. Amidst the complexities of designing \textit{StorySage} to be user-driven (DG 1), support natural conversation (DG 2), and generate an accurate and complete autobiography (DG 3), it is important that users do not perceive the system to be sluggish. Maintaining system responsiveness is essential for keeping users engaged with \textit{StorySage} and enhancing their overall user experience.

\section{StorySage}\label{sec:implementation}
We begin by illustrating \textit{StorySage} through the experience of a hypothetical user, Nick, who is interested in documenting his life story. Nick begins a conversation with \textit{StorySage}, during which the system asks questions about his childhood. As he talks, memories surface—like composing music as a child and meeting his best friend at basketball practice. After a fruitful conversation, Nick concludes the session and indicates a desire to discuss more about his journey into music in a future session. Afterwards, he receives an initial draft of his autobiography, which he reviews and edits ahead of the next session. This flow defines a single \textit{session}, described in Figure~\ref{fig:product_workflow}. Over time, with each conversation, Nick gradually builds a rich autobiography with \textit{StorySage}.

Underlying Nick's experience with \textit{StorySage} is a three-stage system design: \textbf{(1) Interview Session}, \textbf{(2) Biography Writing}, and \textbf{(3) Subsequent Session Preparation}. As shown in Figure~\ref{fig:architecture_overview}, these components perform fundamentally different tasks, though they are inherently connected. This structure naturally lends itself to a multi-agent architecture in which responsibilities are distributed across specialized agents—namely, the Interviewer, Session Scribe, Planner, Section Writer, and Session Coordinator. Although each agent is responsible for executing different tasks, these roles are inherently connected, so the agents rely on shared data structures to coordinate their actions, illustrated in Figure~\ref{fig:architecture_overview}. This modularity is particularly valuable given how developing an autobiography is an iterative process that demands a well-organized system which can adapt and scale as the autobiography evolves over time. In this section, we describe the architecture and implementation of \textit{StorySage}, guided by our design goals in Section~\ref{sec:design_goals}.

\subsection{Interview Session}\label{sec:user_interview} 

At the core of \textit{StorySage} is the interview session—a space for the user to share memories that ultimately form their autobiography. The interview session is led by the \textbf{Interviewer Agent}, who is responsible for facilitating a natural and personalized conversation with the user (DG 2). To foster a sense of intimacy, the Interviewer utilizes a memory bank and question proposal mechanism to ask follow-up questions that align with the user's interests (DG 1). Playing a key role in the question proposal system is the \textbf{Session Scribe Agent} who listens to the conversation in the background and suggests follow-up questions the Interviewer can ask the user. By offloading the responsibility of question generation from the Interviewer—among other responsibilities—\textit{StorySage} ensures a smooth interaction between the user and Interviewer (DG 2). 

Unlike existing work that structures the conversation between user and system around a list of fixed seed questions, \textit{StorySage} is designed with an interface that allows users to steer the conversation by skipping questions or directly suggesting topics they want to discuss. The novelty lies not only in providing users with this level of control, but also in the design of dedicated agents that can dynamically adapt in real time by pivoting the conversation or proposing deeper follow-up questions.

\subsubsection{Interviewer Agent} 
 The Interviewer's sole responsibility is to facilitate a natural and responsive conversation aligned with the user's interests. Therefore, we prompt the Interviewer to propose contextually appropriate questions by monitoring signals of user engagement with the current topic. High-engagement responses tell the Interviewer to ask deeper follow-up questions, while low engagement answers—like unenthusiastic responses or skipped questions—signal the need for a change of topic. To ensure the Interviewer can access a diverse set of questions across different topics and levels of depth, it reads from a dynamic \textbf{session agenda}. The session agenda contains various conversation starters suggested by the Session Coordinator from the previous sessions, as well as deeper follow-up questions that the Session Scribe finds pertinent to the current conversation. The ability to ask follow-up questions from the session agenda reduces the need for synchronous question generation, which can be time-consuming for the Interviewer and introduce latency in the user interface (DG 4). At the same time, when no suitable questions are available, the Interviewer is instructed to propose its own follow-up questions to keep the conversation flowing naturally.

In order to establish a more human-like connection with a user, the Interviewer should have a memory of prior interactions with them. Therefore, inspired by Generative Agents~\citep{park2023generative}, we design the Interviewer to read from two memory modules: a short-term memory that embeds the current session’s dialogue history in the LLM prompt and a long-term \textbf{memory bank}, as shown in Figure~\ref{fig:architecture_overview}. Unlike prior work that embeds a static summary of past conversations directly into the prompt, StorySage’s inter-session memory allows for a more targeted retrieval over a longer horizon. The Interviewer can autonomously invoke a \emph{recall} tool that performs a semantic similarity embedding search over the memory bank to retrieve related memories from prior conversations. This enables the Interviewer to respond thoughtfully in follow-up questions by drawing connection between meaningful memories. For example, in response to the Interviewer's opening question in Figure~\ref{fig:session_scribe_example}, our hypothetical user Nick shares \emph{“I was seven when my mom introduced me to my first instrument.}" The Interviewer can then query the memory bank using phrases such as "\emph{Nick's first instrument}" and "\emph{Nick's mom}" to retrieve memories like \emph{“Nick's first instrument was a piano.”} This retrieval is performed through a similarity search between the query and the stored memories in the memory bank~\citep{sun2023generativeknowledgeselectionknowledgegrounded, li2024corpuslmunifiedlanguagemodel}. The Interviewer can then respond with \emph{“I remember you mentioned that your first instrument was a piano. Did your mother teach you how to play the piano, or did you take lessons?"}

\subsubsection{Session Scribe Agent}

\begin{figure}
\centering
\includegraphics[width=\linewidth]{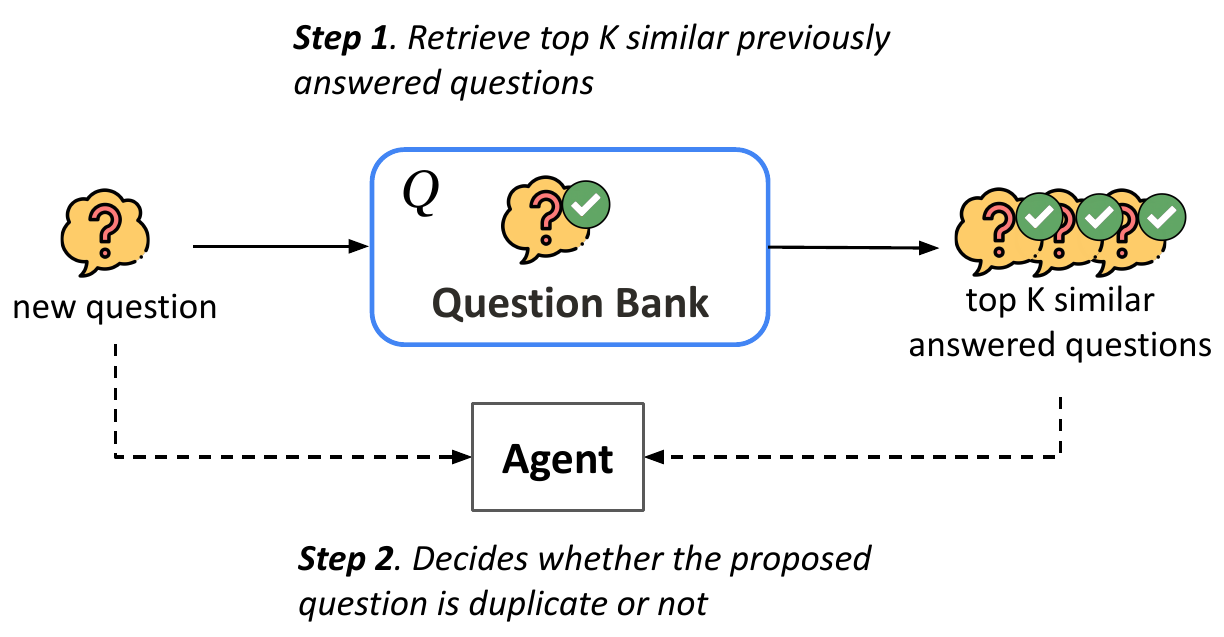}
\caption{
\textbf{Similar Question Detection Mechanism}. \\ This mechanism is utilized by the Session Scribe and Session Coordinator. These agents first (1) retrieve the top K=3 most similar questions from the question bank and then (2) compare the proposed question against the retrieved questions to determine if the proposed question is repetitive.
}
\label{fig:question_repe}
\end{figure}

The Session Scribe acts as an assistant to the Interviewer that listens to the conversation and performs several bookkeeping tasks behind the scenes. It runs concurrently, allowing the user to continue interacting with the system, despite having to handle a number of tasks in the background (DG 4). This design offloads the work of documenting the user's memories and proposing follow-up questions, enabling the Interviewer to focus solely on facilitating a fluid and responsive conversation. As illustrated in Figure~\ref{fig:session_scribe_example}, the Session Scribe updates several data structures in parallel. After hearing Nick share his journey into music, the Session Scribe does the following: 

\begin{enumerate}
\item \textbf{Memory Decomposition:} The Session Scribe decomposes Nick's answer into discrete memories, annotates each with metadata (e.g., event date, location, people involved), and stores them in the memory bank.

\item \textbf{Question Bank Management:} The Session Scribe creates a list of questions that can be implicitly answered from the Nick's response, and adds these to a \textbf{question bank}. This allows future questions to be compared against those previously asked to avoid repetition and maintain natural conversation flow (DG 2).

\item \textbf{Updating Session Agenda:} The Session Scribe records Nick's response and marks the Interviewer's question as answered in the session agenda.

\item \textbf{Follow-Up Question Proposal:} 
The Session Scribe thinks of follow-up questions based on the memory Nick shared. To ensure variety and relevance, we design a mechanism for detecting and filtering repeated follow-up questions.
Each question undergoes a similarity check against the question bank, as illustrated in Figure \ref{fig:question_repe}; only novel questions are added to the current session agenda.
The Session Scribe proposes \textit{fact-gathering questions} to build initial rapport and gather essential context about a memory and \textit{deeper questions} to explore connections between multiple memories and the underlying themes they reveal \cite{jimenez2021prompts, biographicalinterviewing2006}. 

\end{enumerate}

\begin{figure*}[t]
  \centering
  \includegraphics[width=\textwidth]{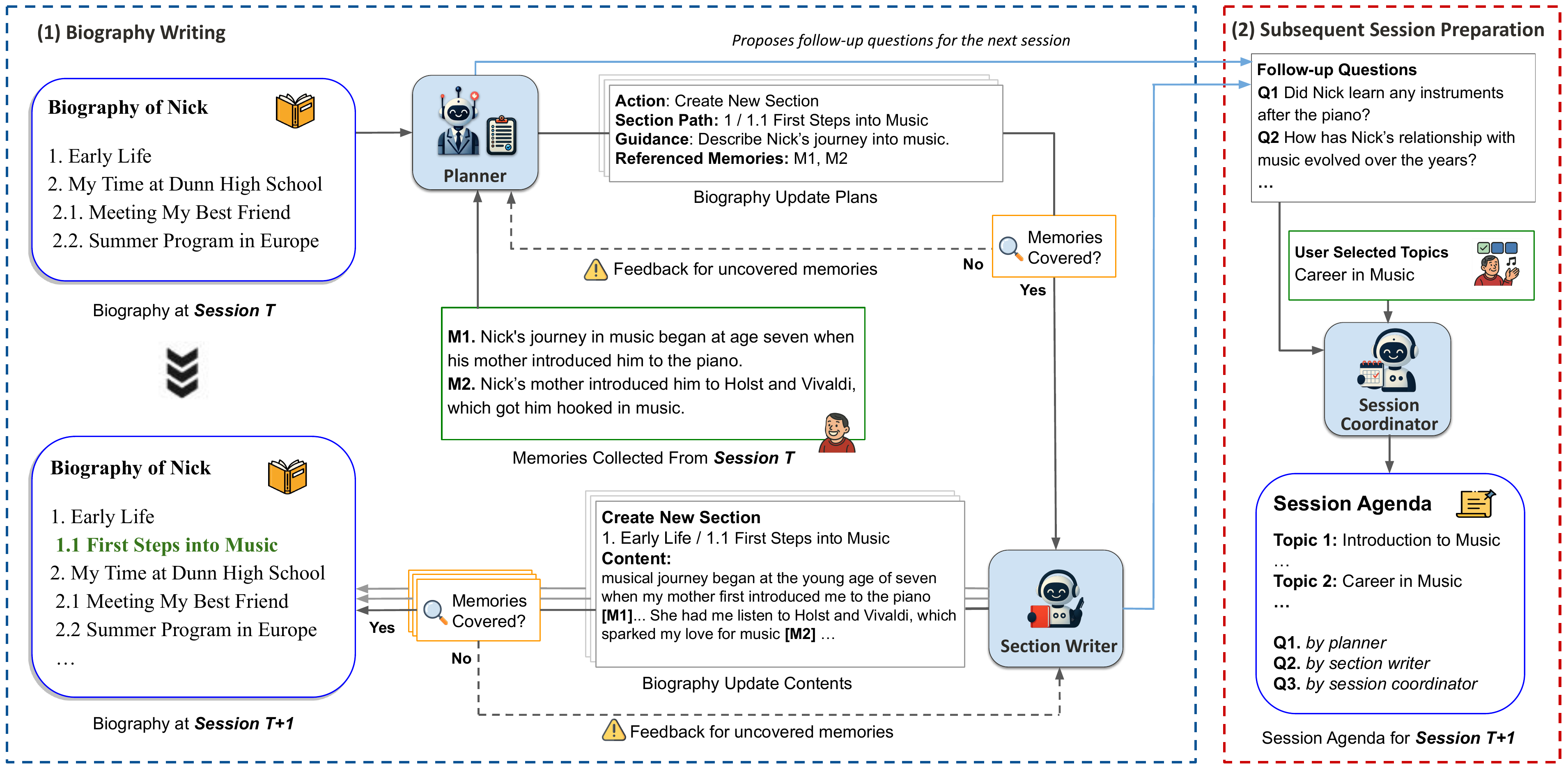}
  \caption{\textbf{Example of the Post-Interview Process}
    \textbf{(1) Biography Writing.} The Planner proposes biography update plans given the existing structure of the biography and the memories collected from the interview. These plans first undergo a memory coverage feedback loop to ensure all memories are integrated into an update plan. The update plans are then given to the Section Writer, who generates narrative content for the autobiography, ensuring that all relevant memories are incorporated into the text. During this process, the Planner and Section Writer each propose follow-up questions. \textbf{(2) Subsequent Session Preparation.} These questions, along with the topics the user selects from the topic selection modal, are given to the Session Coordinator, who prepares the session agenda for the next session.}
  \label{fig:post_interview_process}
\end{figure*}

\subsection{Biography Writing} \label{sec:biography_writing}

The purpose of the biography writing team is to incorporate the user's memories from the interview session into their autobiography. This process is managed by a dedicated writing team comprised of a \textbf{Planner Agent} and \textbf{Section Writer Agent}. As shown in Figure~\ref{fig:post_interview_process}, the Planner first comes up with a set of guidelines for updating the biography using the existing biography structure and the new memories it obtains from the interview session. The Section Writer then executes these plans in parallel, either by adding new sections or revising existing content. The key novelty of our approach lies in decomposing autobiography writing into planning and writing stages, which allows the story structure to evolve flexibly over time, unlike prior work that constrained each session to a predefined chapter. We design these agents to periodically update the autobiography when the number of new memories during the interview session reaches a threshold of 10 unprocessed memories. This prevents a large backlog of memories at the end of the session, which would otherwise prolong writing time (DG 4). After each session, \textit{StorySage} provides users with a draft of their evolving autobiography so they can shape their story in real time by editing the narrative structure and correcting any hallucinations (DG 1). Figure~\ref{fig:example_ui_demo} provides an example of the biography editing interface.

\subsubsection{Planner Agent} 

The Planner follows a structured approach that involves (1) memory grouping, (2) update plan generation, and (3) coverage verification. Figure~\ref{fig:post_interview_process} outlines these in detail. After grouping related memories, the Planner generates an update plan for each group, which is handed off to the Section Writer to add to the autobiography. A feedback loop then ensures that all memories shared by the user are integrated into an update plan, and ultimately the autobiography (DG 3). We use a similar mechanism to ensure the user's edits to the autobiography are properly addressed.

\subsubsection{Section Writer Agent}
The Section Writer is responsible for writing stories for the autobiography using the update plans provided by the Planner.  These update plans contain the relevant memory identifiers and the user's original dialogue that describes these memories.  This focused context window allows the Section Writer to write a narrative that accurately reflects the user’s memories and covers all memories in the update plan (DG 3). The system then writes to different sections of the autobiography concurrently, so users experience minimal delay in receiving their updated autobiography after each session (DG 4).  Like the Planner, the Section Writer is also triggered when users edit the framing of their autobiography.

\subsection{Subsequent Session Preparation}\label{sec:subsequent_session_preparation}

After the user ends the interview session, the biography writing team writes any remaining memories to the autobiography. This marks the beginning of a future session planning phase, a process led by the \textbf{Session Coordinator Agent}. This agent is responsible for preparing the detailed session agenda for the subsequent session with guiding questions that align with the user's interests. To identify these conversational areas, users are presented with a topic selection modal with a list of talking points they expressed interest in from previous sessions (DG 1). This approach helps tackle the challenge of long-form storytelling overlooked in prior work. Instead of relying on static, predefined questions each session that are seemingly disconnected, \textit{StorySage} maintains continuity by constructing future session agendas that are more pertinent to users' conversational interests. Preparation of this agenda concludes the current session and sets the stage for the next session.

\subsubsection{Session Coordinator Agent} 
As illustrated in Figure \ref{fig:post_interview_process}, the Session Coordinator collects a list of follow-up questions from (1) new questions it generates using user-selected topics in the topic selection modal, (2) unanswered questions from the previous session agenda, and (3) follow-up questions proposed from each agent in the biography writing team. The Session Coordinator utilizes the \emph{recall} tool to identify gaps in the memory bank and determine which questions are likely to offer novel insights that can enrich the autobiography. Candidate questions undergo similarity verification against the question bank (Figure~\ref{fig:question_repe}), and redundant questions are either revised or abandoned.

\section{Technical Evaluation}\label{sec:automated_evaluation}

To evaluate \textit{StorySage's} performance over extended use, we conduct a series of experimental studies in which we observe user proxies interact with both \textit{StorySage} and a \textit{Baseline} system over a number of interview sessions. The goal of this study is to assess the basic functionality and behavior of both systems across several key areas prior to user testing, and to evaluate performance when prompted with different underlying language models.

\subsection{Baseline System}\label{sec:baseline_system} 
To understand whether \textit{StorySage}'s multi-agent architecture contributes to better system performance, we construct a \textit{Baseline} system by drawing inspiration from prior work in autobiography writing. We then ablate on features critical to interview quality, user autonomy, and biography writing.


Following GuideLLM \cite{duan2025guidellmexploringllmguidedconversation}, we design the \textit{Baseline} with an Interviewer agent that is equipped with a question outline proposed in "The Life Story Interview" protocol \citep{mcadams2008life}. This outline includes a fixed list of seed questions that ask about reflective life topics, providing strong conversation starters and enabling high-level topic and context navigation. Question proposal is done solely by the Interviewer. It is instructed to think step-by-step \citep{wei2023chainofthoughtpromptingelicitsreasoning} by first selecting a topic from the outline relevant to the user's response; it then formulates a question that suits the current conversational context. To equip the \textit{Baseline} with long-term memory across sessions, we include a summary of prior conversation in the prompt of the Interviewer, consistent with the approach used in GuideLLM. This design allows us to test whether a concurrent multi-agent architecture improves system responsiveness (DG 4) and conversational quality (DG 2) through better question proposal, reduced question repetitiveness, and quicker question proposal time.

In addition to modifying the Interviewer, the \textit{Baseline} system omits the next session topic selection feature, dynamic session agenda, and the biography editing mechanism. For fairness of comparison, we still provide users with their autobiography after each session.
By ablating these features, which provide user autonomy over directing the conversation and narrative flow (DG 1), we can measure how the interactivity and personalization in the design of \textit{StorySage} influences user engagement and overall satisfaction.

In the \textit{Baseline}, we consolidate the Planner and Section Writer into a single Writer agent, thus removing the planning module and iterative feedback loops. To ensure fair comparison and prevent context overload \citep{liu2023lostmiddlelanguagemodels} in the \textit{Baseline}, we chunk the newly collected memories into groups of ten during the interview session. At the end of the session, the Writer sequentially integrates these memory chunks into the autobiography. Moreover, we provide the \textit{Baseline} access to the same biography writing tools as \textit{StorySage}, so the Writer can modify specific sections of the autobiography, rather than having to regenerate the full narrative, ensuring fair latency comparison. The \textit{Baseline} system also follows the same writing guidelines as \textit{StorySage}, which instruct the system to link narrative sentences to memory identifiers and not to hallucinate stories beyond the memories the user provides. By omitting specialized agents and a structured update design from the \textit{Baseline} system, while preserving its writing capabilities, we can compare performance between the \textit{Baseline} and \textit{StorySage} to measure the effect of our multi-agent design on memory coverage and accuracy (DG 3).

\subsection{Experimental Setup}

\subsubsection{User Proxies}\label{sec:simulated_users}
To simulate realistic interview sessions, we design an LLM-powered \textbf{User Agent} tasked with role-playing a human participant \citep{park2023generative, duan2025guidellmexploringllmguidedconversation, zhang-etal-2018-personalizing}. Each user agent is provided with an individual's biography. To encourage the user agent to share a variety of memories, we use a pre-processing step to extract high-level life chapters from their biography. A different chapter is provided as context to user agents at the start of each session, mimicking how real users can choose different life topics to discuss with \textit{StorySage}. We select four Wikipedia biographies of individuals from diverse backgrounds: 
Paul Coates (activist, male) \citep{user_coates}, Jennifer Doudna (biochemist, female) \citep{user_doudna}, Esther Duflo (economist, female) \citep{user_duflo}, and Maggie Rogers (musician, female) \citep{user_roger}. 

\subsubsection{Procedure}
All user proxies interact with both systems for 10 interview sessions, with each session consisting of 20 questions. This is repeated three times per proxy, each time using a different model: GPT-4o \citep{openai2024gpt4ocard}, Gemini-1.5-pro \citep{geminiteam2024unlock}, and DeepSeek-V3 \citep{deepseekaiv32025}. The user proxies are powered by GPT-4o mini \cite{openai2024gpt4omini}, selected for its balance of cost and quality in simulating user behavior.

\subsubsection{Metrics}

We define quantitative metrics to evaluate the basic functionality of \textit{StorySage} and the \textit{Baseline} \cite{duan2025guidellmexploringllmguidedconversation}. 
These metrics were chosen because they provide objective indicators of system performance and reliability, which we find important prior to conducting user testing. 
System performance metrics related to latency are reported in Appendix~\ref{appendix:eval}; here, we focus specifically on biography evaluation metrics:

\begin{itemize}

\item \textbf{Memory Count (\#):} 
The total number of memories collected from all interview sessions.

\item \textbf{Biography Coverage (\%):}
The proportion of memories from the memory bank ($M$) that are referenced in the biography ($B$), calculated as \[\frac{\# \text{ memories covered in } B}{\# \text{ memories stored in } M}\]

\item \textbf{Biography Accuracy (\%):} We use an LLM-as-a-judge approach (details in Appendix~\ref{appendix:groundedness}) to calculate the percentage of total claims in the autobiography that are substantiated by the user's original responses. 

\end{itemize}

\begin{table}[htbp]
\centering
\caption{
Biography evaluation metrics for \textit{StorySage} and \textit{Baseline} across three different underlying models.
Metrics include: \textbf{(Mem.)} total number of memories, \textbf{(Bio. Cov.)} biography coverage, and \textbf{(Bio. Acc.)} biography accuracy.
}
\resizebox{\columnwidth}{!}{%
\setlength{\tabcolsep}{2pt}
\begin{tabular}{llccc}
\toprule
\textbf{Model} & \textbf{System} & \textbf{Mem. (\#)} & \textbf{Bio. Cov. (\%)} & \textbf{Bio. Acc. (\%)} \\
\midrule
\multirow{2}{*}{ Gemini-1.5-pro} &  Baseline & 299 & 86.52 & 96.42 \\
 &  StorySage & 338 & 99.06 & 97.82 \\
\midrule
\multirow{2}{*}{ GPT-4o} &  Baseline & 392 & 61.73 & 98.64 \\
 &  StorySage & 398 & 96.16 & 99.29 \\
\midrule
\multirow{2}{*}{ DeepSeek-V3} &  Baseline & 426 & 77.15 & 99.92 \\
 &  StorySage & 429 & 95.13 & 99.49 \\
\bottomrule
\end{tabular}
}
\label{tab:system_performance_metrics}
\end{table}

\subsection{Results}

Table \ref{tab:system_performance_metrics} presents a performance evaluation of \textit{StorySage} and the \textit{Baseline}. 
To understand how the behavior of each system evolves over time, Figure \ref{fig:simulated_3_3_grid} in Appendix \ref{appendix:eval} illustrates the progression of the key quantitative metrics across multiple interview sessions. \textit{StorySage} extracts more user memories from the conversation compared to the \textit{Baseline} with Gemini-1.5-pro, but a roughly equivalent number with GPT-4o and DeepSeek-V3. Biography coverage is highest with Gemini-1.5-pro, although \textit{StorySage} consistently achieves higher coverage than \textit{Baseline}. Both systems maintain high biography accuracy scores across all models. 

Considering the importance of responsiveness for maintaining a smooth user experience (Appendix \ref{appendix:eval}) and the need for strong memory extraction and coverage, we select GPT-4o as the underlying model for both the \textit{Baseline} and \textit{StorySage}.

\section{User Evaluation}\label{sec:user_evaluation}

While simulations offer useful insights into system performance, they cannot fully capture the holistic user experience, which is central to the evaluation of a user-facing product like \textit{StorySage}. To address this, we conduct a user study framed around answering four \textbf{research questions} (RQ1–RQ4), \\

\begin{enumerate}
    \item User's sense of agency in directing \textit{StorySage}.
    \item Users' perception of the \textit{StorySage's} conversational ability.
    \item Users' satisfaction with \textit{StorySage's} autobiography.
    \item Users' perception of the responsiveness of \textit{StorySage}.
\end{enumerate}
\subsection{Ethics and Research Disclosure} This study was approved by our institution's IRB. All participants provided informed consent prior to their involvement. Participants were compensated at a rate of \$20 per hour. All collected data were treated as confidential and used solely for research purposes \cite{ACM2023ethics}.

\begin{figure*}[t]
  \centering
  \includegraphics[width=\textwidth]{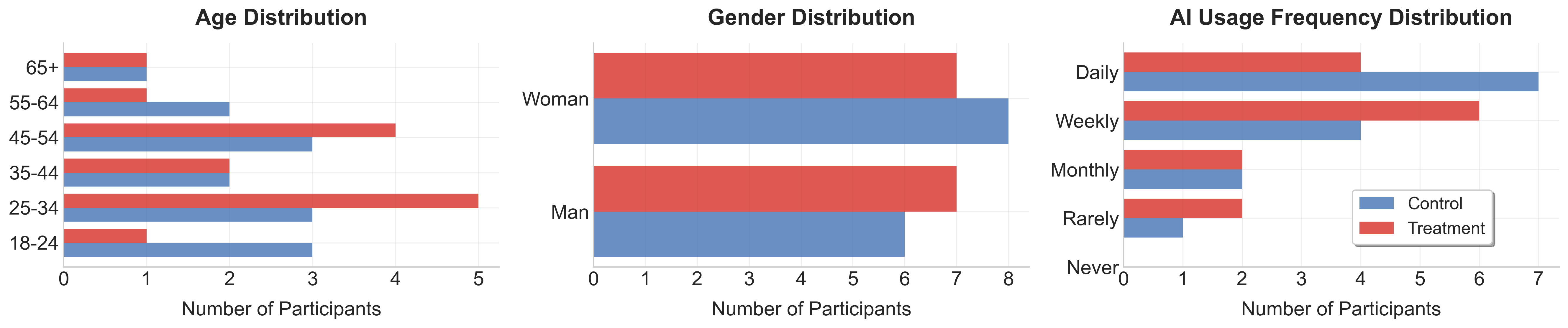}
  \caption{The demographics of participants across the control and treatment group (each $n=14$) in terms of age, gender, and prior AI (ChatGPT, Siri, etc) usage frequency. In aggregate, both groups display similar participant profiles across these three dimensions.}
  \label{fig:demographics_distribution_comparison}
\end{figure*}

\subsection{Experimental Setup} Following standard experimental practice, we assign participants to two equal-sized groups: a control group and treatment group. Participants in the control group assess both the \textit{Baseline} system and \textit{StorySage}, while the treatment group evaluates only \textit{StorySage}. To prevent score contamination, those in the control group evaluate their experience with the \textit{Baseline} prior to engaging with \textit{StorySage}. This approach enables both fair between-group comparisons and within-group analysis to compare how participants in the control group evaluate each system. We primarily rely on results from the between-group analysis, as this design minimizes the risk of carryover, fatigue, and learning effects that could confound results \cite{Budiu2023}. However, the within-group analysis (Appendix~\ref{subsec:within_group}) adds a complementary dimension that enables direct comparison of \textit{StorySage} and \textit{Baseline} within the same group of users and offers deeper insight into individual-level effects \cite{Cuttler2017, Simkus2023, Meuleman2023}.

\subsubsection{Procedure} 
In line with our methodology, participants in the control group spend 45 minutes with each system, while those in the treatment group engage with \textit{StorySage} for 45 minutes. Each 45-minute period is structured into three parts: the showing of a 5-minute introductory video, a 30-minute interaction period, and a 10-minute evaluation period. After giving verbal consent to the IRB-approved study conditions and completing a demographic questionnaire, participants begin watching an introductory video that describes the features of the first system they will interact with. They are then directed to start the first of two 15-minute interviews. Throughout the process, we monitor the interview to ensure a smooth experience, guiding participants to end each interview and read their initial autobiography. This is repeated for a second session, culminating in the participants reading their final autobiography. As a result, on average, participants answered 21.17 questions (\textit{SD} = 8.73) across two 15-minute sessions. Participants then complete a questionnaire, where they provide numerical scores to 7-point Likert-scale questions assessing their experience with the system. These questions are provided in Appendix~\ref{appendix:questionnaire}. Individuals in the control group repeat this process for \textit{StorySage}.

\subsubsection{Participants.}

We recruit 28 participants (15 females, 13 males) from various professional backgrounds from Upwork~\citep{upwork}, explicitly targeting individuals interested in writing their autobiography. Notably, a disproportionately high number of participants come from backgrounds in humanities and computer science, reflecting the common freelance work found on Upwork. We recognize that participants from these backgrounds may offer more critical feedback on AI writing systems, so we stratify participants across gender, age, and professional background to assess how a diverse user base evaluates \textit{StorySage}. These demographics are described in  Figure~\ref{fig:demographics_distribution_comparison} for each experimental group. It is important to note that Upwork's user base tends to skew younger with $82\%$ of participants ranging in age from $18-54$. Participants were asked about their familiarity with AI tools; individuals in control group reported \textit{M} = 4.21, \textit{SD} = 0.97 (on a 5 point scale), while the treatment group reported \textit{M} = 3.86, \textit{SD} = 1.03.

\subsubsection{Data Analysis}
 We conduct statistical significance tests to assess whether participants rate \textit{StorySage} more favorably than the \textit{Baseline} across questionnaire items, both in between-group and within-group settings, reflecting our hypothesis that \textit{StorySage} provides a better user experience across our key design dimensions. Given the ordinal nature of Likert-scale data and the small sample size, we utilize a Wilcoxon rank-sum test \cite{mannwhitneyutest2008} to evaluate our hypothesis in the between-group setting. For our within-group analysis of the control group, we utilize a Wilcoxon signed-rank, a commonly used non-parametric test for comparing paired samples \cite{HOLMES202015}. These tests are applied to each question in the questionnaire, and we report the $p$-values along with the difference in median score, a measure robust to outliers in small sample sizes.

\subsection{Between-Group Analysis}
\begin{figure*}[t]
  \centering
  \includegraphics[width=\textwidth]{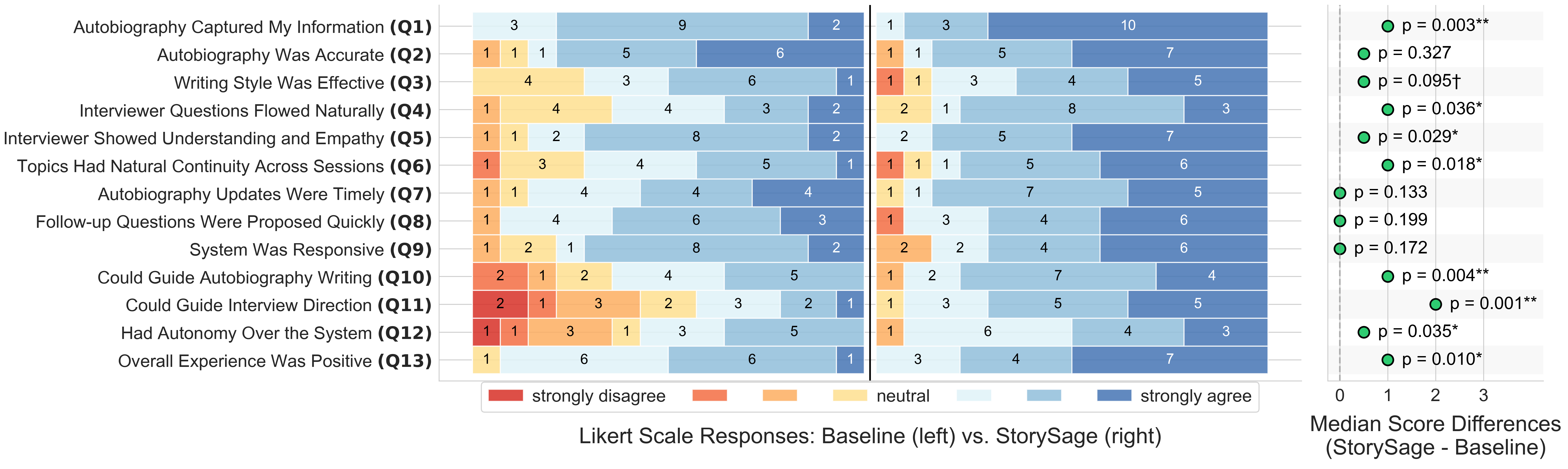}
  \caption{Likert-scale scores from the first system evaluation collected from two independent groups (n=14 per group). The control group rated the \textit{Baseline}, while the treatment group evaluated \textit{StorySage}. The figure presents the median score differences between \emph{StorySage} and the the \textit{Baseline} system for each question. Differences are especially prominent in Q1,3 (autobiography quality), Q4-6 (system conversational ability), and Q10-12 (user autonomy). P-values for the median differences, derived from the Wilcoxon rank-sum test, indicate statistical significance: \textbf{**}$p<0.01$, \textbf{*}$p<0.05$, \textbf{†}$p<0.1$. }
  \label{fig:likert_viz_control_vs_treatment_median}
\end{figure*}

Figure \ref{fig:likert_viz_control_vs_treatment_median} displays a distribution of scores given to the \textit{Baseline} and \textit{StorySage} across each survey item in the between-group setting, where we compare the control group's evaluation of the \textit{Baseline} against the treatment group's evaluation of \textit{StorySage}. Figure~\ref{fig:likert_bar_comparison_control_vs_treatment} provides a bar graph visualization of the scores. The questionnaire consists of 13 Likert-style questions: three questions designed to evaluate each research question, and one final question that assesses participants’ overall experience (see the Appendix \ref{appendix:questionnaire} for question formulation). Overall, users experience a higher level of satisfaction interacting with \textit{StorySage} over the \textit{Baseline} (Q13: median difference = 1, $p$ = 0.01). One participant described how interacting with \textit{StorySage} felt "\textit{like talking to a new friend}" (gender: F, age: 45-54), while another said "\textit{it wrote a better biography for me than I could do myself}" (gender: F, age: 45-54). These responses, along with others, highlight key qualitative themes that we discuss below. 

\subsubsection{System Effect on User Autonomy (Ability to Guide StorySage) [RQ1]} 
\textbf{Participants who interacted with \textit{StorySage} reported a statistically significantly higher level of overall autonomy compared to those who interacted with the \textit{Baseline} model (Q12: median difference = 0.5, p = 0.035).}  \\

\textit{StorySage allows users to better guide the conversation flow during an interview session.} As indicated by Figure \ref{fig:likert_viz_control_vs_treatment_median}, participants reported an ability to more meaningfully guide the interview session (Q11: median difference = 2, $p = 0.001$). This can be attributed to features in \textit{StorySage} that allow users to direct the conversation (e.g., next session topic selection), combined with the Interviewer's ability to respond to subtle user cues—like brief or unanswered responses— with contextually appropriate follow-up questions. Participants reported "\textit{I really liked being able to indicate easily what questions I did not feel like answering}" (gender: F, age: 25-34), and another "\textit{appreciated the option to select topics for future sessions}" (gender: F, age: 45-54). While some participants tried to exert a similar level of control over the \textit{Baseline}, one user "\textit{felt I could not `guide' the questions a certain way, that I was merely responding to the questions}" (gender: F, age: 55-64). Regarding the \textit{Baseline}'s limited ability to change the conversation based on contextual cues, one user noted "\textit{I expected the system to ask follow-up questions, but I don't feel like it ever did}" (gender: M, age: 25-34). Meanwhile, a participant in the treatment group described how \textit{"StorySage picked up on my desire to not continue on a topic from not answering more on a question"} (gender: M, age: 55-64). \\

\textit{StorySage allows users to better guide the writing of their autobiography.} Our study found that the writing process with \textit{StorySage} is significantly more collaborative than that of the \textit{Baseline} system (Q12, median difference = 1, $p = 0.004$). Users attributed this to their ability to edit the biography periodically—a feature that stems from our design choice to update the biography after each session. They appreciated being able to interact with the biography after each session, explaining "\textit{I really like how it was able to quickly start creating the autobiography}" (gender: F, age: 25-34). Many users took advantage of their ability to progressively edit their biography after each session to incorporate their voice and writing style, citing "\textit{the option to edit the product is great}" (gender: F, age: 25-34).

\subsubsection{Perception of System Conversational Ability [RQ2]}
\textbf{Users indicate that \textit{StorySage} facilitates a more natural conversation than the \textit{Baseline} (Q4: median difference = 1, $p$ = 0.036)}. \\

\textit{Participants experienced more natural conversation with \textit{StorySage}, which can be attributed to the Interviewer’s ability to ask more contextually-relevant follow-up questions in an intimate manner.} A participant described how "\textit{StorySage was responsive to my answers and I enjoyed the specificity of the questions...in relation to my previous responses}" (gender: F, age: 18-24). Another explained how "\textit{StorySage picked up on my desire to not continue on a topic by me not answering more on a question}" (gender: M, age: 55-64). We attribute this to \textit{StorySage's} user-centric design that encourages the Interviewer to analyze engagement before deciding whether to shift topics or probe deeper. This approach differs from the conversational style of the \textit{Baseline}, which uses predefined seed questions similar to those found in existing work~\citep{duan2025guidellmexploringllmguidedconversation}. A user explained how "\textit{the conversation [with the Baseline] felt less like a conversation and more like a rigid set of prompts the continue regardless of response}" (gender: M, age: 25-34). We also find that it is more difficult for the \textit{Baseline} to pick up on the emotion in a user's answer and respond appropriately (Q5: median difference = 0.5, $p$ = 0.029). For example, a participant in the control group reports that the \textit{Baseline} "\textit{didn't include some of the emotions I had stated I was feeling during those particular moments the Interviewer asked for}" (gender: F, age: 45-54). However, user feedback also suggests that \textit{StorySage} occasionally asks too many follow-up questions focused around a singular topic. When this happens, participants change the conversation topic, which further underscores the importance of providing them greater control and autonomy. For example, a participant explained "\textit{I had to force StorySage to change topics...because it asked too many questions about a certain topic}" (gender: F, age: 18-24). We elaborate on this limitation in Section~\ref{sec:limitations}. \\

\textit{StorySage is better at maintaining a natural continuity between successive interview sessions} (Q6: median difference = 1, $p$ = 0.018). This can be attributed to \textit{StorySage's} next session planning module. A user explicitly mentioned how they "\textit{appreciated the option to select topics [for the next session]}" (gender: F, age: 45-54), while others appreciated starting the second session with topics of their choice when prompted by the Interviewer's open-ended opening question. On the other hand, multiple participants in the control group noticed that the \textit{Baseline} would tend to ask repeated questions across the two sessions. This is likely due to the \textit{Baseline} system’s weaker inter-session memory, which relies on a summary of past conversations~\citep{duan2025guidellmexploringllmguidedconversation}. In contrast, \textit{StorySage} has a question bank and employs a similarity-based verification mechanism to prevent redundant questions. "\textit{During the second session ... [the Baseline] did ask very similar questions that it had asked before. This repetitiveness felt a bit unnatural}" (gender: F, age: 18-24).

\subsubsection{User Satisfaction with the Autobiography [RQ3]}
\textbf{Users were moderately more satisfied with StorySage's autobiography.} \\

\textit{\textit{StorySage} provides a more complete autobiography that captures users’ memories more comprehensively than the \textit{Baseline}} (Q1: median difference = 1, $p$ = 0.003). This outcome is likely due to the design of the Planner and Section Writer, who work together to ensure high biography memory coverage through focused context windows and content verification loops. However, we also observed that content was occasionally repeated within \textit{StorySage’s} autobiographies, suggesting a potential overuse of the feedback mechanism. We discuss this limitation in more detail in Section~\ref{sec:limitations}. For example, a participant noted that "\textit{there was some repetition of information about my teenage years}" (gender: F, age: 65+). Nonetheless, participants in the treatment generally expressed satisfaction with the completeness of their autobiographies, as reflected in the number of high scores given to Q1. We notice a similar positive sentiment among participants in the control group, with a participant reporting "\textit{I appreciated that [the Baseline] didn't write every little aside and joke I made}" (gender: F, age: 45-54). However, high memory coverage tends not to scale with session length in the \textit{Baseline}, as illustrated by our long-running simulations in Table \ref{tab:system_performance_metrics}. We attribute this difference to the absence of the multi-agent design and verification mechanism in the \textit{Baseline}, which enables \textit{StorySage} to maintain high coverage as session length increases.

Table~\ref{tab:bio_stats} presents a high-level, quantitative comparison of the autobiographies generated by \textit{StorySage} and the \textit{Baseline} for users in the treatment and control groups, respectively. On average, autobiographies produced by \textit{StorySage} are nearly twice as long as those of the \textit{Baseline}, cover nearly three times as many memories, and include a greater number of sections. These differences suggest that \textit{StorySage} may be able to write more comprehensive and organized narratives, likely due to its planning and verification mechanisms. However, these results cannot be attributed solely to the biography writing team, as the increase in biography length may also stem from \textit{StorySage’s} ability to ask less repetitive questions and foster more natural conversation, which helps elicit more memories. \\

\textit{We find no significant difference in the accuracy of the biography produced by both systems} (Q2: median difference = 0.5, $p$ = 0.327). High evaluation scores suggest that both systems are capable of generating a narrative that accurately reflects the content of the user's memories. One user shared "\textit{[StorySage] did a good job understanding what I was saying and keeping everything in the correct context}" (gender: F, age: 55-64). We hypothesize this stems from our design choice to provide the user's original memory—as they expressed it during the conversation—to the both systems. Notably, as shown in Table~\ref{tab:bio_stats}, \textit{StorySage} biographies are substantially longer than those produced by the \textit{Baseline}. Despite this increase in length, participants did not report reduced accuracy, suggesting that the additional content is not the result of hallucination or irrelevant elaboration, but rather reflects more complete coverage of the user’s shared experiences. However, in a few cases, participants noted that both systems tended to overstate the emotional tone of certain memories, which some users perceive as a form of inaccuracy. \\

\textit{Users prefer the writing style of the autobiography produced by StorySage somewhat more than the writing style of the Baseline} (Q3: median difference = 0.5, $p$= 0.095). A participant in the treatment group explained "\textit{it wrote a better biography for me than I could do myself}" (gender: F, age: 45-54). With a Section Writer solely focused on generating narrative content from the Planner's update plans, \textit{StorySage} can spend more resources to produce an eloquently written autobiography \cite{yao2019plan, simon2022tattletale, wang2023improvingpacinglongformstory}. However, qualitative feedback suggests that participants across both groups prefer different writing styles—some wanted a professional tone, while others preferred a narrative that retained their voice. For example, one user who spoke to the system with short, concise sentences felt "\textit{the tone of writing didn't feel like me}" (gender: F, age: 25-34). This is closely related to the system's tendency to exaggerate the emotional tone of the narrative, similar to the last feedback point we mention. We expand on these limitations of \textit{StorySage} in Section~\ref{sec:limitations}.

\begin{table}[H]
\centering
\small
\begin{tabular}{lcccc}
\toprule
System & Words (W) & Memories & Sections (S) & W/S \\
\midrule
StorySage & 1064.14 & 33.36 & 5.64 & 199.05 \\
(Treatment) & {\footnotesize $\pm$189.14} & {\footnotesize $\pm$8.97} & {\footnotesize $\pm$1.86} & {\footnotesize $\pm$43.44} \\
\midrule
Baseline & 557.36 & 13.79 & 4.00 & 152.28 \\
(Control) & {\footnotesize $\pm$210.09} & {\footnotesize $\pm$5.94} & {\footnotesize $\pm$1.80} & {\footnotesize $\pm$52.90} \\
\bottomrule
\end{tabular}

\vspace{1em}

\caption{Biography statistics for \emph{StorySage} and \emph{Baseline}. Metrics represent mean $\pm$ standard deviation across users in each group.}
\label{tab:bio_stats}
\end{table}

\subsubsection{Perception of System Responsiveness [RQ4]}
\textbf{Participant felts that \textit{StorySage} and the \textit{Baseline} system exhibited similar levels of responsiveness (Q9: median difference = 0, $p$ = 0.172).} \\

\textit{Users did not report a statistically significantly faster question proposal time with StorySage.}
Despite \textit{StorySage}'s empirically faster question proposal speed (as shown in Section~\ref{subsec:app_system_latency}), participants rated both systems similarly with a median score of 6 on a 7-point Likert scale (Q8: median difference = 0, $p = 0.199$). This does not imply that concurrency is ineffective; instead, it enables \textit{StorySage} to handle background processes that enhance conversational quality without incurring additional latency compared to the \textit{Baseline}. Moreover, analysis of the outlying scores indicated that some participants considered the audio playback delay after the follow-up question appeared on the screen, when evaluating the systems' question proposal latency.  This delay occurs due to the conversion delay of the text-to-speech function, which was identical in both systems. While one participant reported that "\textit{the voice asking the questions felt delayed compared to when the text was displayed}" (gender: F, age: 25-34), another reported "\textit{the system was amazingly responsive}" (gender: F, age: 65+). Our analysis indicates that differences in latency on the order of a few seconds can be difficult to detect in user studies due to the subjective perception of time.  \\

\textit{Participants did not notice a significantly faster autobiography generation time with StorySage after each session.} Users evaluated the speed of autobiography writing, and in aggregate, they found that new memories were incorporated in both autobiographies at similar speeds (Q7: median difference = 0, $p = 0.133$). Despite the computational overhead of a multi-step planning and writing workflow, the concurrent design of \textit{StorySage} maintains a comparable biography generation time as the \textit{Baseline}. Notably, because biography generation begins before the topic selection modal is shown to users, part of the true latency is masked, making \textit{StorySage} appear slightly faster. However, for fairness of comparison, we report objective latency measurements in our experimental simulations (Appendix~\ref{subsec:app_system_latency}).
Analyzing qualitative feedback, we find that the evaluation results are subjective. One user reported, "\textit{I really like how [StorySage] was able to quickly create the autobiography within 10 minutes or less}" (gender: F, age: 25-34), though some users may still prefer a faster turnaround.

\section{Discussions}\label{sec:discussions}

In this section, we reflect on key insights from our study, discuss limitations of the current system, and outline directions for future work, including important ethical considerations.

\subsection{Design Implication}

\subsubsection{Human-in-the-Loop Design}
Supporting human-in-the-loop interaction is important when designing AI systems for long-form, creative tasks because it keeps users actively engaged in both the generative and decision-making aspects of the process~\citep{dhillon2024shaping}. Moreover, our study suggested that fully AI-led writing may overlook the value some individuals place on the mental effort of writing, particularly in work that represents their personal identity. Moreover, hallucinations produced by these systems may lead to a loss of trust and misrepresentation of personal memories, further highlighting the need for human involvement. In the case of \emph{StorySage}, participants felt a greater sense of autonomy in contributing to their autobiography because the system offered them an ability to steer the conversation, revise their biography periodically, and influence the narrative direction across sessions. Crucially, the system was designed to be responsive to their input, which reinforced their sense of control and transformed the interaction into a more collaborative experience. These findings suggest that creative AI systems should be designed to continuously involve the user. This is especially important for creative tasks that unfold over longer periods of time, where ongoing human involvement helps users stay engaged and ensures they retain influence over the system's outputs. Rather than operating fully autonomously, these systems should strike a balance where both the system and the user actively contribute to shaping the outcome~\citep{dhillon2024shaping, buschek2021impact}.

\subsubsection{Modularity} Modular design is useful for building AI systems that needs to manage several moving parts in a coordinated way~\citep{wooldridge2009multiagent}. In the case of \emph{StorySage}, a multi-agent framework made it possible to split distinct responsibilities across five specialized agents. This separation of concerns allowed each agent to operate within a well-scoped context, enabling better organization that improved both the conversational flow and the writing process~\citep{huot2024agents} without sacrificing system responsiveness. Creative AI systems designed to support work developed over longer periods can benefit from modular architectures, which can provide both scalability and responsiveness, while allowing for flexibility in how different components of the system can evolve over time.

\subsection{Future Work and Limitations}\label{sec:limitations}
\paragraph{Conversational navigation.} 
Our qualitative analysis reveals that people engage in conversations differently, making it challenging to build a system that feels tailored to everyone~\citep{ha2024clochat}. To address this, \textit{StorySage} allows users to guide the conversation and ask follow-up questions that align with their interests. While this helps personalize the conversation, it does not fully address the problem. At its core, the Interviewer is powered by an LLM with a manually crafted prompt, and given context that includes the session agenda, chat history, and memory recall tools. However, it is not fine-tuned to navigate conversations. Therefore, the Interviewer occasionally becomes overly focused on a single topic.  While some users enjoy this style of conversation, others find it too deep. Future work can improve the Interviewer's navigation capabilities by incorporating feedback mechanisms that prevent the questions from becoming unnecessarily detailed or even fine-tuning the base language model on the conversational dialogue~\citep{villa2024comparative}. Similarly, the Interviewer's engagement recognition capabilities can be improved by leveraging finetuned models, such as EmoLlama~\citep{Liu_2024, duan2025guidellmexploringllmguidedconversation}.

\paragraph{High memory coverage and repetition.} Although our experimental simulation show that \emph{StorySage} is able to achieve near-perfect biography memory coverage, user feedback highlights an issue with memory repetition across the autobiography. We hypothesize that this arises from the system's memory coverage verification loops, which may contribute to redundancy. Future work can involve users more directly during the planning phase, allowing them to select which memories they want reflected in their narrative. This approach can also help preserve their voice in the autobiography and provide greater autonomy during the writing process.

\paragraph{Biography reconstruction.}
Supporting content reconstruction is important for creating longer, coherent autobiographies. Although users do not explicitly raise this concern, our current design does not support automatic reorganization; rather, we provide a mechanism for users to manually reorganize their biography during the editing process. Automatic reconstruction becomes valuable for longer narratives, as manual reconstruction is both difficult and time-consuming. Future research can explore strategies for automatically merging related content and generating multiple outlines that users can choose from. Additionally, future work can more closely align with conventional practices of professional biographers, who hold multiple conversations before outlining a biography. In this sense, the overall planning process can benefit from accumulating a larger memory bank prior to structuring.

\paragraph{Semantic hallucination.} While neither \textit{StorySage} nor the \textit{Baseline} introduces inaccurate memories in the autobiography, users note that the system occasionally describes memories using adjectives or emotions they had not shared during the interview. This effect appears to stem from the Section Writer's tendency to write the autobiography with a positive tone, even when provided with the user's framing of the memory. We attribute this behavior to the model's post-training process, where the RLHF encourages positively framed writing. To help re-incorporate tone back in the narrative, \textit{StorySage} allows users to edit their biography, although we realize this does not solve the deeper problem. Future work can offer multiple phrasings of the autobiography, potentially generated by a model fine-tuned for biography writing. Additionally, such a system can present writing samples to users throughout the writing process to better learn their preferred narrative style.

\paragraph{Longitudinal evaluation.} Our primary findings are drawn from a user study in which participants engage with \emph{StorySage} across two 15-minute conversational sessions (30 minutes total). To examine the scalability of our system design, we conducted a simulation study using LLM-based user proxies over 10 sessions, with 20 conversational rounds each. While this simulation demonstrated that \textit{StorySage} can function effectively over longer sessions and multiple rounds of interaction, it does not capture how real users perceive the system over extended use. Future research should evaluate how users evaluate \textit{StorySage} across our key design dimensions when interacting with the system over extended periods of time.

\paragraph{System design.} We acknowledge that our design lacks rigorous ablation studies to validate each architectural choice. 
Additionally, we recognize that other frameworks may also be viable; as such, the system should be viewed as one possible approach that manages the interview and autobiography writing processes. Key design choices, including conversation navigation, memory management, and planning strategies, may have room for improvement. As with many LLM-based systems, the performance of our system is sensitive to variations in prompts and the underlying language model. Future work can measure the effect of individual design choices and optimize system components, prompts, and models.

\subsection{Ethical Risks and Considerations}
A system like \emph{StorySage}, which is powered by LLMs, raises several ethical considerations, particularly related to data privacy, narrative accuracy, model bias, and the potential for over-reliance.

\paragraph{Data privacy and confidentiality.} Autobiographical writing involves sharing personal and sensitive memories, which raises concerns about privacy and confidentiality ~\citep{Yao_2024, yan2024protectingdataprivacylarge, alkamli2024understanding}. Sending these narratives to third-party LLM providers can increase the risk of exposing sensitive data~\citep{li2024human}. Although participants in our study were clearly informed about the data handling protocols, broader deployment of such systems should consider using on-premises open-source models to reduce their reliance on external services and provide stronger guarantees around data protection~\citep{feretzakis2024trustworthy}.

\paragraph{Narrative distortion and model bias.} LLMs are known to reflect and amplify biases present in their training data~\citep{lucy-bamman-2021-gender, hu2025generative}, which can lead to plausible yet inaccurate narratives. In the context of autobiography writing, research has shown that this can distort the narrative and introduce unintended perspectives that compromise the authenticity of a user's life story~\citep{luther2024teaming, Huang_2025, kim2024mindfuldiary, hwang202480, jakesch2023co}. To maintain the accuracy and integrity of these narratives, robust content verification measures and human oversight are essential ~\citep{schulz2024impact}.

\paragraph{Over-reliance.} Although \emph{StorySage} incorporates several features designed to provide users with a greater sense of control, individuals may opt to under-engage with the system. While convenient, this over-reliance diminishes the cognitive and emotional value of the writing process~\citep{li2024valuebenefitsconcernsgenerative, dhillon2024shaping, kim2024diarymate, zhai2024effects, passi2022overreliance, lu2021human}. 
User involvement is critical in maintaining a co-creative dynamic that ensures the autobiography reflects the user's voice.

\section{Conclusion}

In this paper, we introduced \textit{StorySage}, a conversational assistant designed to support collaborative autobiography writing through human-AI co-creation. Our multi-agent framework decomposes this complex writing task into three distinct components—interviewing, biography generation, and session planning—each handled by a specialized agent. Rather than operating in a fully autonomous fashion, \textit{StorySage} is built to both \textit{guide} the user and \textit{adapt} to their input, allowing for a more engaging and collaborative writing experience. Evaluation results show that \textit{StorySage} significantly improves users' sense of autonomy, fosters more natural conversation flow, and makes meaningful progress towards producing complete and accurate autobiographies—all while demonstrating responsiveness. More broadly, \textit{StorySage} illustrates how multi-agent conversational writing systems can advance storytelling and digital legacy preservation, enriching both personal and shared narratives.

\section{Acknowledgment}

This work was supported in part by the Air Force Office of Scientific Research (AFOSR) under Grant
FA9550-23-1-0251 and in part by the Office of Naval Research under Grant N00014-24-1-2164.

\bibliographystyle{ACM-Reference-Format}
\bibliography{references}

\input{appendix}

\end{document}

%% file: appendix.tex
\appendix
\onecolumn

\section{User Evaluation}

\subsection{User Testing Feedback Form}\label{appendix:questionnaire}

Below is the questionnaire provided to participants to assess their experience interacting with \textit{StorySage} and the \textit{Baseline}. Participants provided ratings on a 7-point Likert scale (1 = lowest, 7 = highest), elaborated on their experiences, and consented regarding publication of their generated content.

\begin{tcolorbox}[colback=green!10, colframe=green!40!black, title=User Testing Feedback Form]
\textbf{What is your name?} (Text response)

\vspace{0.2cm}
\textbf{On a scale of 1--7:}
\begin{enumerate}
\item[(Q1)] How well did the autobiography capture the information you shared during your interview session?
\item[(Q2)] How frequently did you notice instances where the autobiography included information that did not accurately reflect what you shared during your interview session?
\item[(Q3)] How do you rate the writing style of the autobiography?
\item[(Q4)] How natural did you find the overall flow of the interviewer’s questions?
\item[(Q5)] How well did the interviewer demonstrate understanding and empathy in response to your answers?
\item[(Q6)] How much of a natural continuity did you feel across the topics discussed in the first and second sessions?
\item[(Q7)] How quickly were updates to the autobiography provided after each session?
\item[(Q8)] How quickly were follow-up questions proposed during the interview session?
\item[(Q9)] How responsive did the system feel when interacting with it?
\item[(Q10)] How much control did you feel you had in influencing the writing of your autobiography?
\item[(Q11)] How much control did you feel you had in guiding the interview’s direction, both during the interview session and in deciding which topics would be covered in future sessions?
\item[(Q12)] How much autonomy did you feel you had over the system?
\item[(Q13)] Rate your overall experience with the system.
\end{enumerate}

\vspace{0.2cm}
\textbf{Please elaborate on your answer to the above question.} (Long text answer)

\vspace{0.2cm}
\textbf{Do we have your consent to publish parts of your generated autobiography or conversational history?} (Yes / No)
\end{tcolorbox}

\newpage
\subsection{Between-Group Analyses}

\begin{figure}[H]
  \centering
  \includegraphics[width=0.9\textwidth]{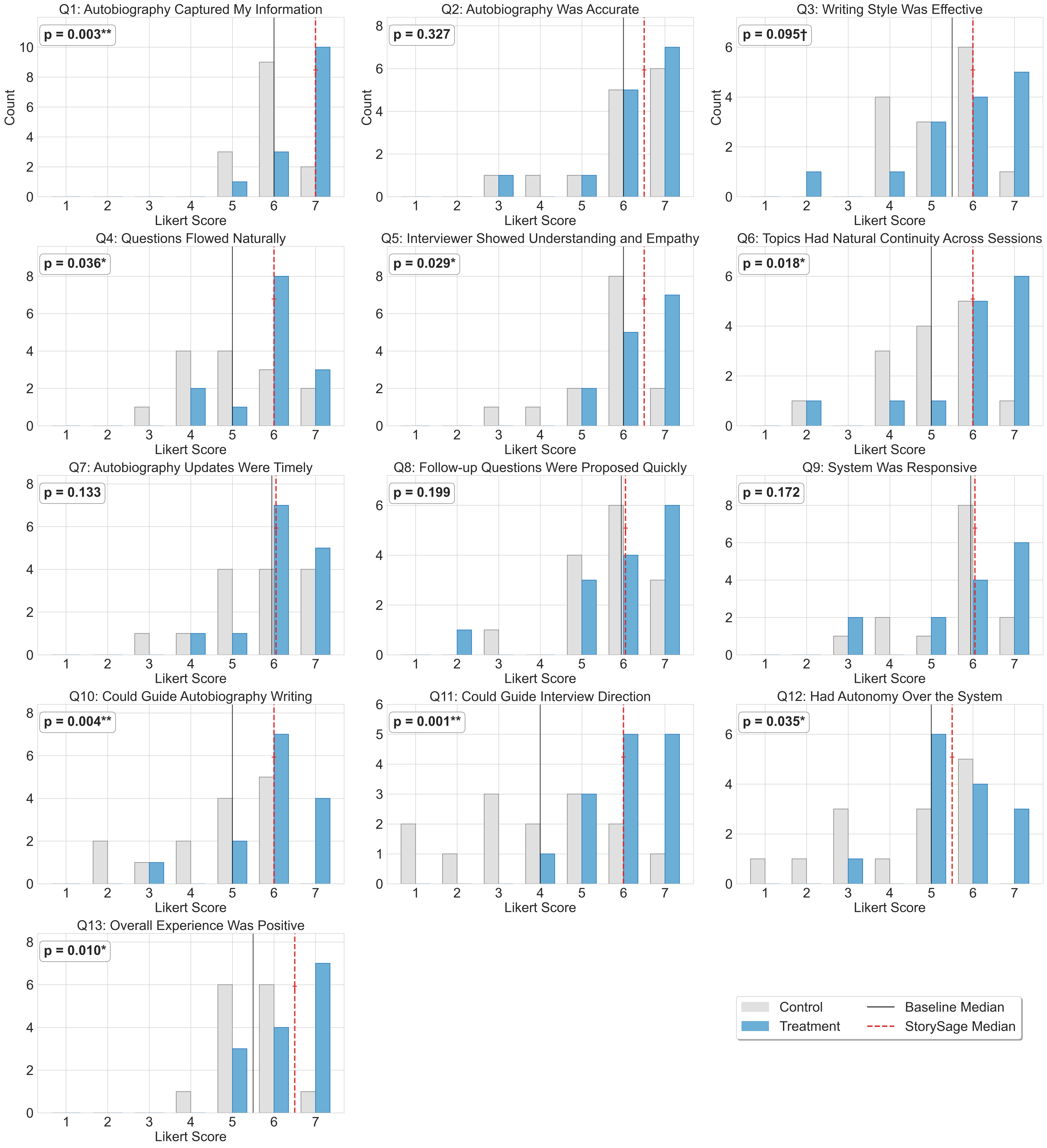}
  \caption{Likert-scale scores from the first system evaluation collected from two independent groups (n=14 per group). The control group rated the \textit{Baseline} system, while the treatment group evaluated \textit{StorySage}. The figure presents the median score for \textit{StorySage} and the \textit{Baseline} for each question. P-values for the median differences, derived from the Wilcoxon rank-sum test, indicate statistical significance: \textbf{**}$p<0.01$, \textbf{*}$p<0.05$, \textbf{†}$p<0.1$. }
  \label{fig:likert_bar_comparison_control_vs_treatment}
\end{figure}

\newpage
\subsection{Within-Group Analyses}\label{subsec:within_group}

\begin{figure*}[tp]  
    \centering
    \includegraphics[width=\textwidth]{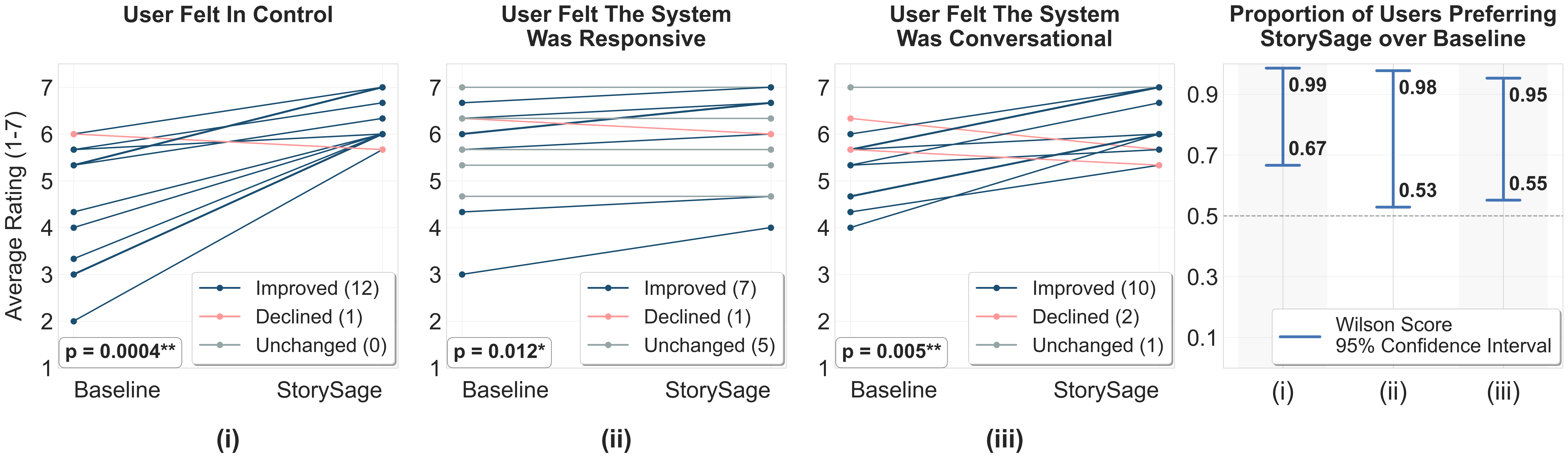}
    \caption{Evolution of Likert-scale ratings from control group participants (n=13) who first evaluate the \textit{Baseline} and then assess \textit{StorySage}. The ratings are aggregated to reflect participants' perception of \textbf{(i)} human autonomy [DG 1], \textbf{(ii)} system responsiveness [DG 4], and \textbf{(iii)} conversational ability [DG 2], based on the average scores of responses for questions 10–12, 7–9, and 4–6, respectively. The fourth subplot shows confidence intervals of the proportion of users who strictly prefer \textit{StorySage} across all three dimensions (i–iii), computed using a Wilson score interval. P-values in (i), (ii), and (iii) are derived from the Wilcoxon signed-rank test and indicate statistical significance: \textbf{**}$p<0.01$, \textbf{*}$p<0.05$, \textbf{†}$p<0.1$.}

    \label{fig:spaghetti_plot_category_changes}
\end{figure*}
Following our experimental design, participants in the control group interact with \textit{StorySage} after evaluating the \textit{Baseline}. The advantage of within-group tests is that participants can directly compare the two systems, which may increase their sensitivity to differences by allowing them to exercise comparative judgment \cite{charness2012experimental}. Comparing their questionnaire responses then provides us insights into how the same users perceived differences between the systems. Figure~\ref{fig:spaghetti_plot_category_changes} presents the aggregated results that compute the average scores of the respondents, following the common practice of combining multiple Likert-scale items that assess the same construct into a single composite score \cite{Boone2012,Gliem2003}. This approach is further justified by the high correlation we observed between questionnaire responses. One user is excluded from our analysis due to a technical error during their session. 

To estimate the true proportion of users who prefer \textit{StorySage} across the dimensions we identify in our research questions, we construct 95\% confidence intervals using the Wilson Score method \citep{WikipediaBinomialProportionCI} and find that a statistically significant majority of users strictly prefer the autonomy, responsiveness, and Interviewer conversational ability of \textit{StorySage} (RQ1-RQ3). Interestingly, we observe a statistically significant preference for the responsiveness of \textit{StorySage} in the within-group study, but not in the between-group study—particularly in faster question proposal. We hypothesize that this is because participants in the within-group setting evaluate \textit{StorySage} after interacting with the \textit{Baseline}, giving them a more concrete basis of latency in an existing system for comparison. Moreover, this result somewhat aligns with the latency difference we highlight in our experimental simulations (Section~\ref{sec:automated_evaluation}). At the same time, it is important to acknowledge that these findings may be influenced by potential carryover effects, including participant fatigue or a bias toward rating the second system more favorably after exposure to a better system.

In total, eight participants in the within-group study preferred \textit{StorySage}, two felt indifferently, and three preferred the \textit{Baseline}. Participants who preferred \textit{StorySage} described the "\textit{interviewing [style] was a lot more in depth and informative}" (gender: F, age: 45–54), and "\textit{[StorySage] felt like more of a conversation than the [Baseline]}" (gender: M, age: 35–44). \textit{StorySage offered much better control on topics} (gender: M, age: 55-64) and "\textit{showcased a pretty great writing style!}" (gender: F, age: 25-34). One participant who preferred the \textit{Baseline} system explained that \textit{StorySage} "\textit{was not repetitive but wanted all the nuances [of my memories]}" (gender: M, age: 25-34), and another described how "\textit{information [was presented] less accurately [in the autobiography]}" (gender: F, age: 45-54). We discuss these points of feedback as well as other limitations in the following section.

\begin{figure}[h!]
  \centering
  \includegraphics[width=0.9\textwidth]{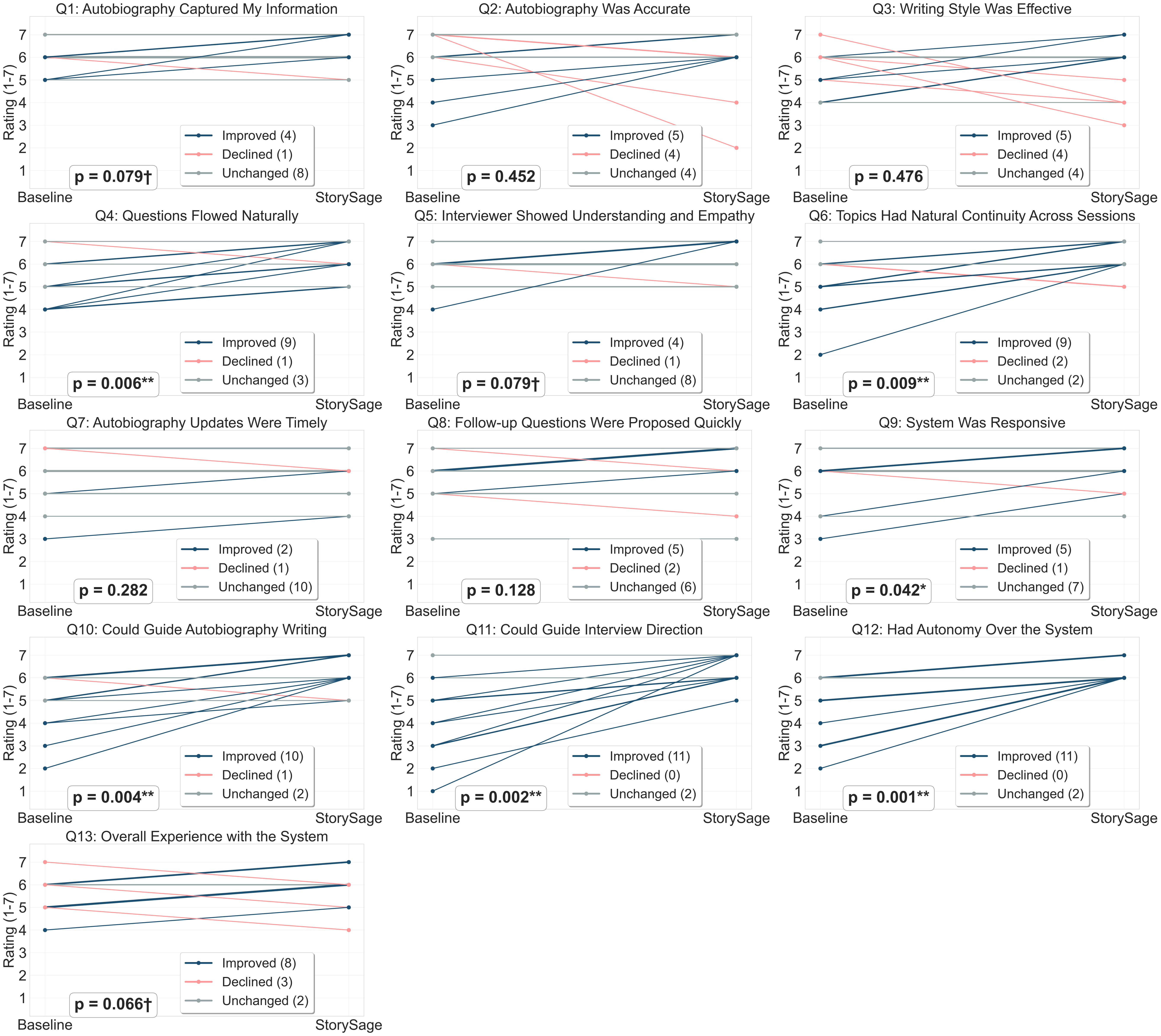}
  \caption{Evolution of Likert-scale ratings from control group participants (n=13) who first evaluate the \emph{Baseline} and then assess \emph{StorySage}. Each plot corresponds to one of 13 questionnaire questions. P-values are derived from the Wilcoxon signed-rank test and indicate statistical significance: \textbf{**}$p<0.01$, \textbf{*}$p<0.05$, \textbf{†}$p<0.1$.}

  \label{fig:spaghetti_plot_user_changes}
\end{figure}

Figure~\ref{fig:spaghetti_plot_user_changes} illustrates changes in questionnaire responses from participants in the control group who evaluated both the \textit{Baseline} and \textit{StorySage}. Some individual questionnaire items, especially those related to system responsiveness (Q8 to Q10), show little change. However, when we average each user's responses across the dimension of each research question, we observe an overall improvement. This suggests that although question-level differences may not reach statistical significance, aggregate patterns may reflect more favorably of \textit{StorySage}.

\newpage
\section{Implementation Details}

\subsection{Shared Data Structures}\label{sec:shared_data_structures}
To facilitate coherent interactions among agents and meaningful user engagement, we implemented four structured data models shared across the multi-agent system:

\begin{itemize}
    \item \textbf{Memory Bank:} A structured repository maintained by the Session Scribe that stores memory entities shared by the user. 
    Each memory is annotated with rich metadata including date, location, associated individuals, and emotional context.
    Memory entities are systematically linked to the originating user response, thereby maintaining provenance and contextual integrity.
    Additionally, each memory includes semantic embeddings to support efficient similarity-based retrieval, enhancing agents' ability to recall relevant content precisely. 
    
    \item \textbf{Session Agenda:} 
    The session agenda provides structure for each interview session, guiding the Interviewer’s progression through the conversation.
    Prepared by the Session Coordinator, the session agenda includes a summary of prior sessions and a set of proposed follow-up questions based on the user’s past responses and selected topics. During the session, the Session Scribe documents user responses and generates follow-up questions in real time. This helps the Interviewer maintain coherence and adapt to new topics as they emerge. 
    
    \item \textbf{Question Bank:} The question bank is a cumulative bank maintained by the Session Scribe that captures all questions asked across sessions. These include questions asked both explicitly and implicitly. Unlike the session agenda, which focuses on the current session, the question bank offers a long-term view of the interaction history. It helps the Session Scribe and Session Coordinator avoid repetition and ask varied questions.
    
    \item \textbf{Biography:} 
    The biography is structured as a tree with chapters, sections, and subsections, offering a clear and organized representation of the user’s life story. The Planner updates the structure as new content emerges, while the Section Writer adds narrative text. This structure allows users to review and refine the biography across multiple sessions. It remains accessible to the user at all times.
    
\end{itemize}

\subsection{Workflow of Session Scribe}

\begin{figure}[H]
  \centering
  \includegraphics[width=0.95\textwidth]{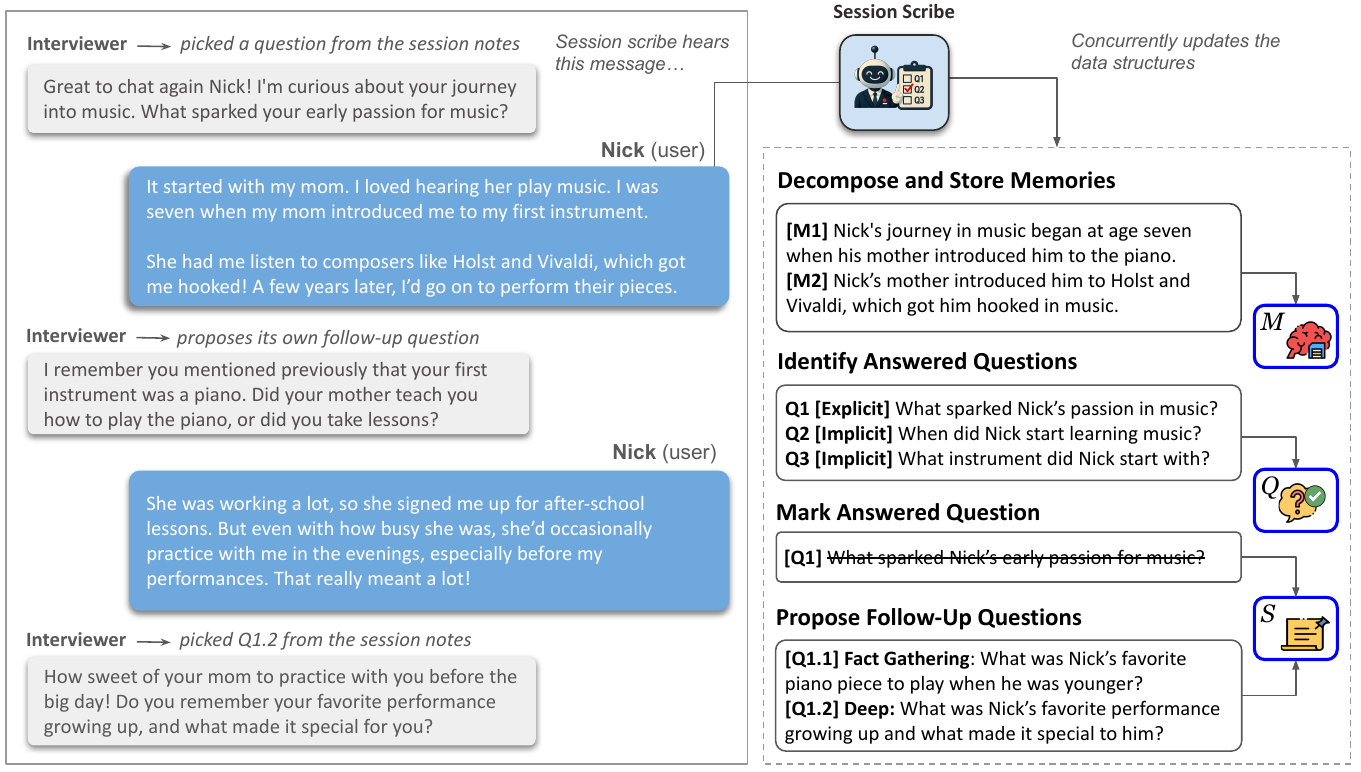}
  \caption{Workflow of the Session Scribe during the interview session. The diagram illustrates how the Session Scribe interfaces with the shared data structures, including the memory bank, question bank, and session agenda.}
  \label{fig:session_scribe_example}
\end{figure}

Figure~\ref{fig:session_scribe_example} displays the workflow of the Session Scribe as it processes information from interview conversations. The diagram shows how the system extracts and organizes user information into the memory bank, question bank, and session agenda. The example shown here is a follow-up to the example presented in Figure~\ref{fig:product_workflow}

\newpage

\subsection{From User's Response to Data Structures}

\begin{figure}[H]
\centering
\includegraphics[width=0.65\linewidth]{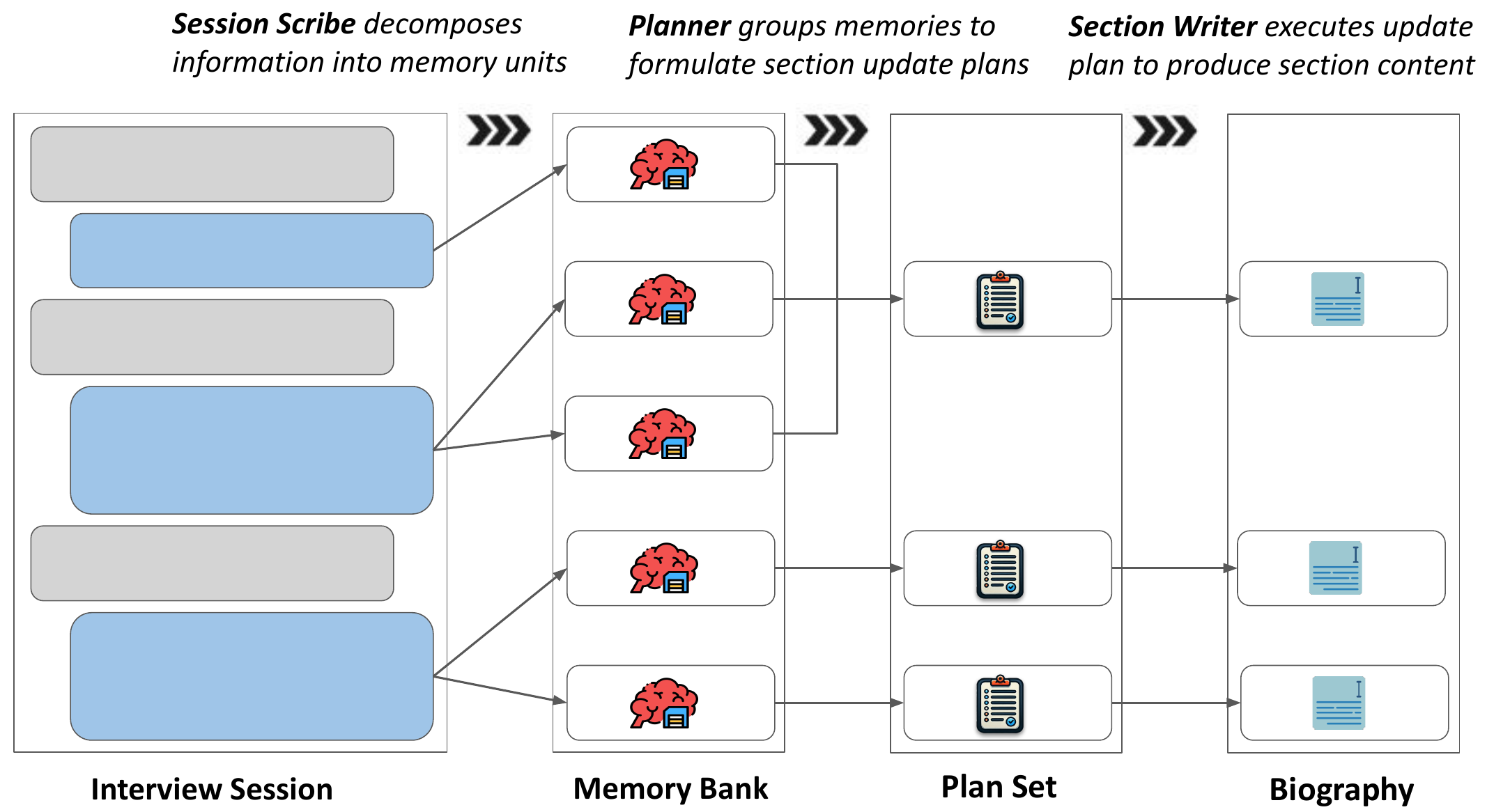}
\caption{\textbf{Flow of the information from the user's chat into the biography.} The Session Scribe decomposes users' responses into memory units. The Planner then groups these memories to formulate biography update plans, which the Section Writer translates into biography content. This three-stage pipeline ensures systematic processing from user voice to structured biographical content.}
\label{fig:data_flow}
\end{figure}

Figure~\ref{fig:data_flow} illustrates how the user conversation data flows through the system’s data structures to finally reflect in the biography.

\subsection{Example of UI process}\label{subsec:editing_ui}

\begin{figure}[H]
\centering
\includegraphics[width=0.95\linewidth]{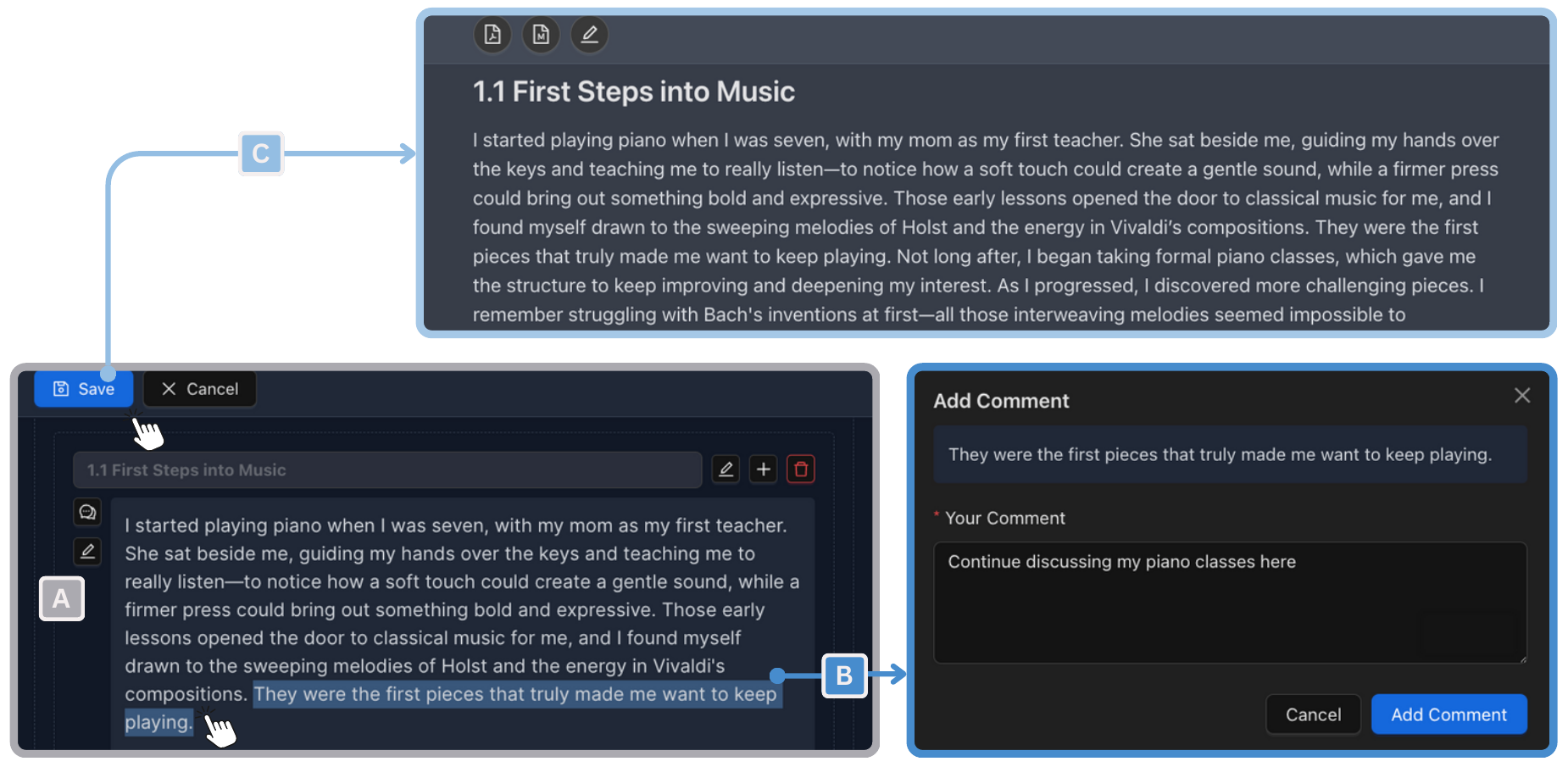}
\caption{Example of biography editing interface workflow. \textbf{(A)} The sequence begins when a user accesses the edit page. \textbf{(B)} Upon text selection, a comment dialog appears, allowing annotation of specific content. \textbf{(C)} After submitting comments and saving changes, the user can review their updated biography. This interactive process facilitates targeted feedback and automated content refinement. Note that users are also given the option to directly edit their narrative.}
\label{fig:example_ui_demo}
\end{figure}

Figure \ref{fig:example_ui_demo} illustrates the biography editing interface workflow, demonstrating the system's interactive revision process.

\newpage
\subsection{Biography Accuracy Evaluation Methodology}\label{appendix:groundedness}

We assess the accuracy of each biography section individually and compute an average score across all sections as the final biography-wide accuracy score. In cases where a biography section is too long to evaluate in a single pass, we use a continuation mechanism in which the model receives its previous outputs as context, allowing it to maintain coherence and complete the accuracy assessment across the full section. For the following section, we refer to "accuracy" and "groundedness" interchangeably.

\begin{tcolorbox}[colback=yellow!5, colframe=yellow!40!black, breakable, title=Section Accuracy (Groundedness) Score]
You are an expert at evaluating if biographical text is grounded in source memories.

Given a biography section and its source memories, evaluate if the biographical text is substantiated by the memories. Return a score between 0-100, indicating what percentage of the biographical content is substantiated by the memories.

Let's define groundedness mathematically:

\begin{enumerate}
\item Let \(S = \{s_1, s_2, \ldots, s_n\}\) be the set of atomic information units from source memories.
\item Let \(B = \{b_1, b_2, \ldots, b_m\}\) be the set of atomic claims/statements in the biography section.
\item Define \(B_{\text{substantiated}} = \{b_i \in B \mid b_i \text{ can be derived from or substantiated by any } s_j \in S\}\).
\item The groundedness score is calculated as:
\[ \text{Groundedness Score} = \frac{|B_{\text{substantiated}}|}{|B|} \times 100 \]
\end{enumerate}

==========================================

\textbf{Biography Section to Evaluate:} 
\begin{verbatim}
{biography_section_content}
\end{verbatim}

\textbf{Source Memories:} 
\begin{verbatim}
{list_of_memories}
\end{verbatim}

==========================================

To calculate the groundedness score, follows these steps in your thinking process.

Step 1: Decompose source memories into atomic information units

- Source Memory 1: [List atomic information units]

- Source Memory 2: [List atomic information units]

...

Step 2: Decompose biography section into atomic claims/statements

- Claim 1: [Statement]

- Claim 2: [Statement]

...

Step 3: Evaluate each claim against source information

- Claim 1: [Substantiated/Unsubstantiated] - [Reasoning]

- Claim 2: [Substantiated/Unsubstantiated] - [Reasoning]

...

Step 4: Calculate groundedness score

- Total claims: [Number]

- Substantiated claims: [Number]

- Groundedness score: [Calculation] = [Final percentage]

\end{tcolorbox}

\newpage

\section{Extended Technical Evaluation}\label{appendix:eval}

\subsection{System Latency}\label{subsec:app_system_latency}

    This section provides a detailed examination of additional technical metrics to assess the computational performance of \textit{StorySage} and the \textit{Baseline} system. In particular, we report latency measurements collected in our experimental setting across three different underlying models, as a means of benchmarking each system’s responsiveness (DG 4).

\begin{itemize}
\item \textbf{Question Proposal Latency (sec).
The time required for the Interviewer to propose a follow-up question.}

\item \textbf{Autobiography Update Latency (sec):
The time required for \textit{StorySage} to generate the autobiography after the end of a session.}
\end{itemize}

\begin{table}[htbp]
\centering
\caption{
System latency metrics (in seconds) for \textit{StorySage} and \textit{Baseline} across different models. All metrics are averaged across 10 sessions and 4 simulated users.
}
\begin{tabular}{llcc}
\toprule
\textbf{Model} & \textbf{System} & \textbf{Q. Lat.} & \textbf{Bio. Lat.} \\
\midrule
\multirow{2}{*}{Gemini-1.5-pro} & Baseline & 6.68 & 90.35 \\
 & StorySage & 3.07 & 39.52 \\
\midrule
\multirow{2}{*}{GPT-4o} & Baseline & 5.19 & 78.98 \\
 & StorySage & 3.16 & 52.72 \\
\midrule
\multirow{2}{*}{DeepSeek-V3} & Baseline & 8.76 & 206.40 \\
 & StorySage & 4.19 & 167.43 \\
\bottomrule
\end{tabular}
\label{tab:latency_metrics}
\end{table}

Across all language models in in Table \ref{tab:latency_metrics}, \textit{StorySage} consistently proposes follow-up questions 2–4 seconds faster than the \textit{Baseline}. This demonstrates that, despite the added complexity of \textit{StorySage}’s conversational architecture, the system achieves stronger computational performance through faster response times—highlighting the benefits of parallelization. However, \textit{StorySage} is slower at biography generation when DeepSeek-V3 and GPT-4o are used as the underlying model, but both systems are equally fast with Gemini-1.5-pro. GPT-4o offers the fastest performance overall in question and biography generation time. 

\subsection{Progression of Biography Metrics}

\begin{figure*}[t]
    \centering
    \begin{tabular}{ccc}        
        \vspace{-0.4cm}
        
        \begin{subfigure}[b]{\textwidth}
            \centering
            \includegraphics[width=\linewidth]{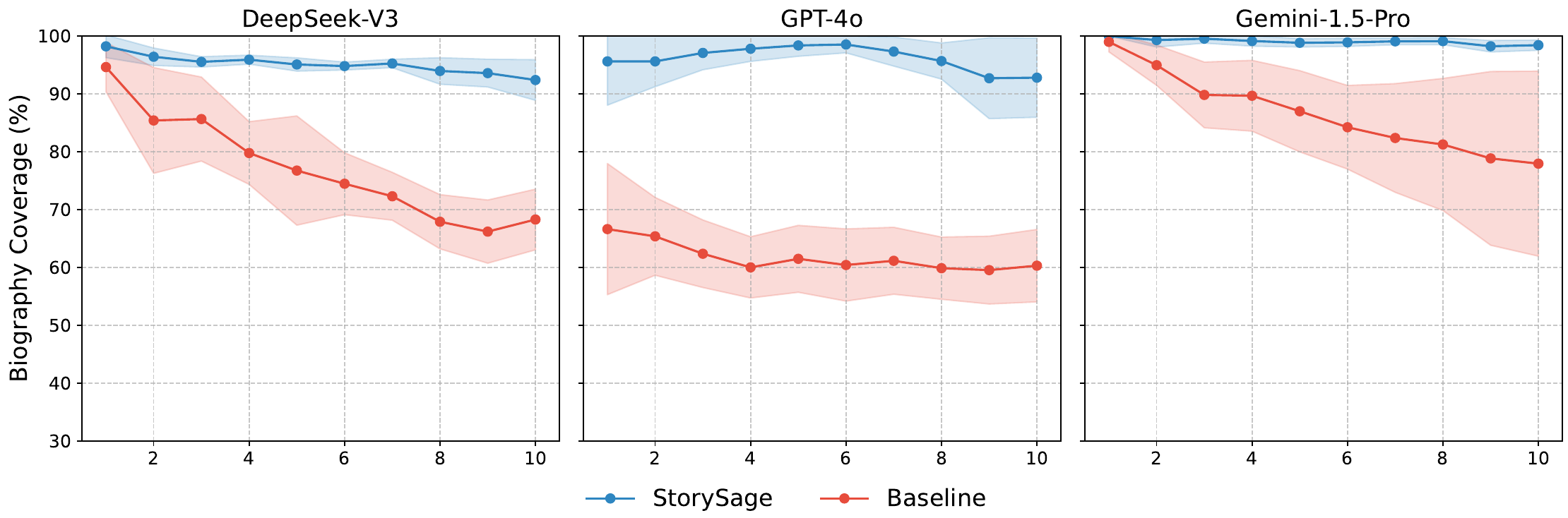}
            \label{fig:completeness-comparison}
        \end{subfigure} \\
        
        \vspace{-0.5cm}
        
        \begin{subfigure}[b]{\textwidth}
            \centering
            \includegraphics[width=\linewidth]{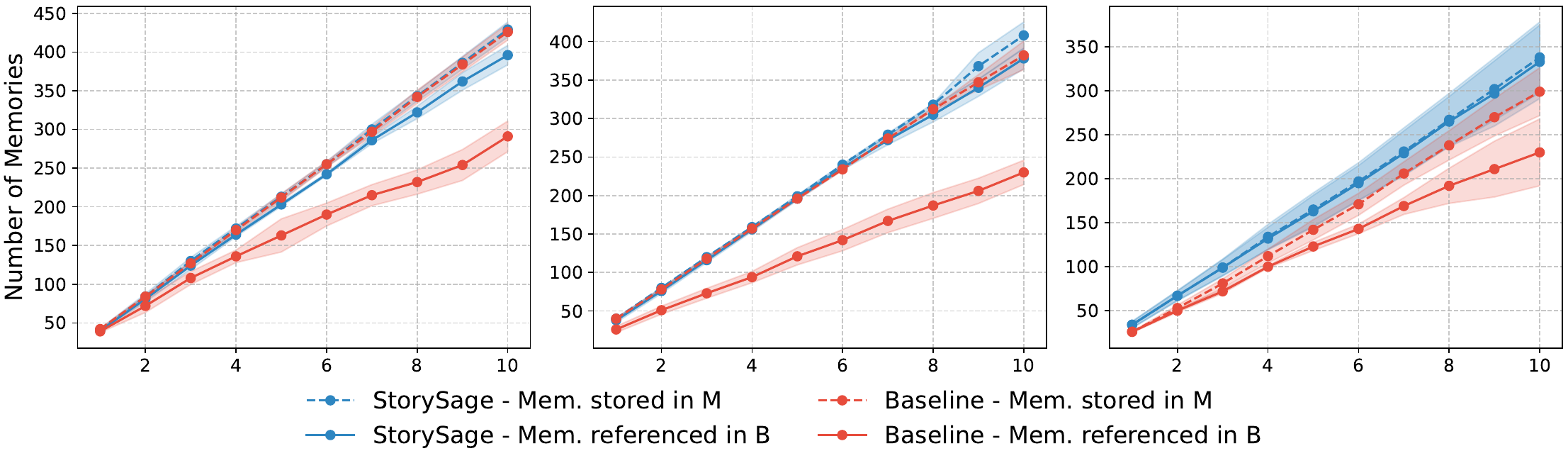}
            \label{fig:memory-counts-comparison}
        \end{subfigure} \\
    \end{tabular}
    
    \caption{
    Performance comparison metrics between \emph{StorySage} and \textit{Baseline} across three underlying language models (DeepSeek-V3, GPT-4o, and Gemini-1.5-Pro). Lines represent average performance across four simulated user agents, with the X-axis indicating session number (1 to 10). Across all models, \emph{StorySage} consistently outperforms the \textit{Baseline}. \textit{Biography Coverage} (top) reflects \emph{StorySage}'s ability to maintain high memory coverage across many sessions, while \textit{Number of Memories} (bottom) reveal both systems ability to extract many memories. In the bottom plots, dashed lines indicate total memories stored (M), and solid lines represent those referenced in the biography (B).
    }
    \label{fig:simulated_3_3_grid}
\end{figure*}

Figure \ref{fig:simulated_3_3_grid} illustrates the progression of key quantitative metrics across multiple interview sessions. The results demonstrate that StorySage's modular approach consistently outperforms the Baseline's single-agent writing strategy, even when the latter leverages advanced long-context models like Gemini-1.5-Pro. This performance gap highlights a fundamental limitation of monolithic approaches: despite having access to extensive context windows, single-agent systems struggle to effectively process and integrate the growing volume of conversational memories into coherent biographies.

Particularly notable is the stark difference in biography coverage (top row). While StorySage maintains consistently high coverage (>90\%) across all models and sessions, the Baseline system shows significant degradation as the number of sessions increases. This degradation is most pronounced with GPT-4o, where the Baseline's coverage drops below 65\%, but is evident even with Gemini-1.5-Pro's enhanced context handling capabilities, achieving only around 80\% when it comes to the 10th session. These results underscore that merely increasing a model's context window is insufficient; the architectural choice of modular memory management and biography generation proves crucial for maintaining comprehensive life narratives.

